\documentclass[runningheads]{llncs}
\usepackage{graphicx}
\usepackage{tikz}
\usepackage{comment}
\usepackage{amsmath,amssymb}
\usepackage{color}
\usepackage{subcaption}
\usepackage{booktabs}
\usepackage{algorithmic}
\usepackage[ruled,linesnumbered]{algorithm2e}
\usepackage{hyperref}
\hypersetup{
  colorlinks,
  linkcolor={black},
  citecolor={black},
  urlcolor={orange!70!red}
}

\usepackage[accsupp]{axessibility}

\newcommand{\mypara}[1]{\smallskip\noindent{\bf {#1}.}}

\newcommand{\ShadowDataset}{\mathcal{D}_{\textit{shadow}}}

\newcommand{\TargetTrain}{\mathcal{D}_{\textit{target}}^{\textit{train}}}
\newcommand{\TargetTest}{\mathcal{D}_{\textit{target}}^{\textit{test}}}
\newcommand{\ShadowTrain}{\mathcal{D}_{\textit{shadow}}^{\textit{train}}}
\newcommand{\ShadowTest}{\mathcal{D}_{\textit{shadow}}^{\textit{test}}}
\newcommand{\TargetModel}{\mathcal{T}}
\newcommand{\ShadowModel}{\mathcal{S}}
\newcommand{\AttackModel}{\mathcal{A}}
\newcommand{\Knowledge}{\mathcal{K}}
\begin{document}

\title{Semi-Leak: Membership Inference Attacks Against Semi-supervised Learning}

\titlerunning{Semi-Leak}

\author{Xinlei He\inst{1} \and
Hongbin Liu\inst{2} \and
Neil Zhenqiang Gong\inst{2} \and
Yang Zhang\inst{1}}

\authorrunning{He et al.}

\institute{CISPA Helmholz Center for Information Security \and
Duke University
}

\maketitle

% ----------------------------------------------------
\begin{abstract}
% ----------------------------------------------------

Semi-supervised learning (SSL) leverages both labeled and unlabeled data to train machine learning (ML) models. 
State-of-the-art SSL methods can achieve comparable performance to supervised learning by leveraging much fewer labeled data.
However, most existing works focus on improving the performance of SSL. 
In this work, we take a different angle by studying the training data privacy of SSL.
Specifically, we propose the first data augmentation-based membership inference attacks against ML models trained by SSL.
Given a data sample and the black-box access to a model, the goal of membership inference attack is to determine whether the data sample belongs to the training dataset of the model.
Our evaluation shows that the proposed attack can consistently outperform existing membership inference attacks and achieves the best performance against the model trained by SSL.
Moreover, we uncover that the reason for membership leakage in SSL is different from the commonly believed one in supervised learning, i.e., overfitting (the gap between training and testing accuracy).
We observe that the SSL model is well generalized to the testing data (with almost 0 overfitting) but ``memorizes'' the training data by giving a more confident prediction regardless of its correctness.
We also explore early stopping as a countermeasure to prevent membership inference attacks against SSL. 
The results show that early stopping can mitigate the membership inference attack, but with the cost of model's utility degradation.\footnote{Our code is available at \url{https://github.com/xinleihe/Semi-Leak}.}

\keywords{Membership Inference Attack, Semi-Supervised Learning}

% ----------------------------------------------------
\end{abstract}
% ----------------------------------------------------

% ----------------------------------------------------
\section{Introduction}
\label{section:introduction}
% ----------------------------------------------------

Machine learning (ML) has made tremendous progress in the past decade.
One of the key reasons for the great success of ML models can be credited to the large-scale labeled data.
However, such labeled datasets are often hard to collect as they rely on human annotations and expertise in the specific domain.
Meanwhile, unlabeled datasets are easy to obtain.
To better leverage the unlabeled data, semi-supervised learning (SSL) has been proposed.
Concretely, SSL uses a small set of labeled data and a large set of unlabeled data to jointly train the ML model.
In recent years, SSL shows its effectiveness on different tasks by leveraging much fewer labeled data~\cite{SBCZZRCKL20,XDHLL20,ZWHWWOS21}.
For instance, by only using 250 labeled samples, FlexMatch~\cite{ZWHWWOS21} can achieve about 95\% accuracy on CIFAR10.

Different from supervised learning where every data sample is treated equally in the training procedure, 
SSL takes different ways to handle the labeled and unlabeled data samples during the training.
Concretely, the state-of-the-art SSL methods~\cite{SBCZZRCKL20,XDHLL20,ZWHWWOS21} leverage weak augmentation to the labeled samples and trains them in a supervised manner.
For each unlabeled sample, it would generate a weakly-augmented view and a strongly-augmented view (by weak and strong augmentations), and the goal is to leverage the model's prediction probability (referred to as prediction or posteriors) of the weakly-augmented view to guide the training of the strongly-augmented view of the sample.
Instead of directly using the posteriors as a ``soft'' label, those SSL methods switch the posteriors into a ``sharpen''~\cite{XDHLL20} or ``hard'' label~\cite{SBCZZRCKL20,ZWHWWOS21}.
Note that the sample is not used to train the model until the highest probability of the prediction on the weakly-augmented view exceeds a pre-defined threshold $\tau$. 
In this way, the model trained by SSL can gradually learn more accurate predictions.

Despite being powerful, ML models are shown to be vulnerable to various privacy attacks~\cite{FJR15,SSSS17,SS20}, represented by membership inference attacks~\cite{SSSS17,SZHBFB19,NSH18,SM21}.
The goal of membership inference attack is to determine whether a data sample is used to train a target ML model.
Successful membership inference attacks can raise privacy concerns as they may reveal sensitive information of people.
For instance, if an ML model is trained on the data for people with a certain sensitive attribute (e.g., diseases), identifying the person in the training dataset directly reveals this individual's sensitive attribute.
So far, most of the efforts on membership inference attacks concentrate on models trained by supervised learning.
Also, there are some exploratory researches investigating the privacy risks in self-supervised learning~\cite{LJQG21,HZ21}.
However, in SSL, the labeled and unlabeled samples are treated differently during the training.
It is important to quantify whether this unique training paradigm would lead to different privacy risks for labeled and unlabeled samples.
Also, as the different augmented views instead of the original samples are used to train the model, we are curious whether a more effective membership inference attack mechanism can be proposed against SSL.
To be best of our knowledge, this is largely unexplored.

In this work, we fill the gap by proposing the first data augmentation-based membership inference attack method against SSL.
A key advantage for SSL is that it only needs a small amount of labeled data and leverages the unlabeled data itself to guide the training.
Concretely, for the labeled data, the model is trained in a supervised manner.
For the unlabeled data, SSL leverage the data itself as the supervision.
In particular, for each unlabeled training sample, a weakly augmented and a strongly augmented views will be fed into the target model and the training objective is to minimize the distance of the model's prediction on these two views.
Our proposed data augmentation-based attack is based on the intuition that the model's prediction of these two views should be more similar if the sample belongs to the model's training set.

We conduct our evaluation on three SSL methods (FixMatch, FlexMatch, and UDA) and three commonly used SSL datasets (SVHN, CIFAR10, and CIFAR100).
Our empirical results show that our proposed attack can consistently outperform baseline attacks and reaches the best performance.
For instance, for FixMatch trained on CIFAR10 with 500 labeled samples, our attack achieves 0.780 AUC while the best baseline attack only has 0.722 AUC.
This indicates that our attack can better unleash the membership information in SSL.

Moreover, we find that, unlike supervised learning where the membership leakage can be credited to the overfitting nature of the model~\cite{SSSS17,SZHBFB19} (i.e., the model predicts the training data more accurately than the testing data), models trained by SSL methods are well generalized and have almost no overfitting but still suffer high membership inference risk.
Our analysis reveals that the model indeed ``memorizes'' the training data, but such memorization does not present as a more accurate prediction, but a more confident prediction.
We show that the prediction entropy distribution of members and non-members has a large gap in models trained by SSL (measured by Jason-Shannon (JS) Distance).

\mypara{Contributions}
(1) We are the first to study the privacy risk of SSL through the lens of membership inference attacks and we propose a data augmentation-based attack that is tailored to SSL methods.
(2) We conduct extensive experiments on SVHN, CIFAR10, and CIFAR100 datasets. Our results show that our proposed attack outperforms baseline attacks that are extended from existing works to SSL settings.
(3) We show that the effectiveness of membership inference attacks against SSL is not credited to the model's overfitting level but credited to the model prediction's distinguishable entropy distributions for members and non-members (measured by Jason-Shannon Distance).
(4) We study an early-stopping-based defense against our proposed attack. We show that this defense can decrease the attack AUC of our attack but sacrifice the testing accuracy of the trained models.

% ----------------------------------------------------
\section{Preliminary and Related Work}
\label{section:preliminaries}
% ----------------------------------------------------

% ----------------------------------------------------
\subsection{Semi-Supervised Learning}
% ----------------------------------------------------

Semi-supervised learning (SSL)~\cite{L13,MMKI19,BCGPOR19,SBCZZRCKL20,XDHLL20,ZWHWWOS21} aims to train accurate models via exploiting a large amount of unlabeled data when the labeled data is scarce. 
In this paper, we focus on the vision domain since most advanced SSL methods are designed for it. 
Generally speaking, state-of-the-art SSL techniques~\cite{SBCZZRCKL20,XDHLL20,ZWHWWOS21} produce ``pseudo labels'' for the unlabeled samples when the model's predictions are confident enough based on pre-defined threshold strategies. 
For example, Lee~\cite{L13} first proposed to produce the class label that has the highest confidence score output by the classifier for unlabeled samples during training. 
After assigning pseudo labels to unlabeled samples, they can train classifiers in a supervised fashion with labeled and unlabeled samples. 
Recently, FixMatch~\cite{SBCZZRCKL20} achieves state-of-the-art classification accuracy via assigning the strongly augmented unlabeled samples with the pseudo labels produced from the corresponding weakly augmented samples when the highest confidence score exceeds a certain threshold.
While UDA~\cite{XDHLL20} was proposed to treat the classifier's ``sharpen'' output confidence scores as the `pseudo labels' rather than one class label. 
Similar to FixMatch, UDA trains strongly augmented unlabeled samples with the pseudo labels produced from the corresponding weakly augmented samples. 
FlexMatch~\cite{ZWHWWOS21} updates FixMatch by introducing the curriculum learning-based method to flexibly adjust the threshold for different classes during the training. 
Existing studies on SSL mainly focus on how to improve the performance, however, we are the first to show that state-of-the-art SSL methods are vulnerable to our tailored membership inference attacks, which exploit the strong/weak data augmentations used by state-of-the-art SSL methods.

% ----------------------------------------------------
\subsection{Membership Inference Attacks}
% ----------------------------------------------------

Membership inference attack~\cite{SSSS17,SZHBFB19,NSH18,PTC18,CYZF20,HRSF20,ROF20,HYYBGC21,LJQG21,HZ21,SM21,HWWBSZ21,LZ21,WGCS21,CCNSTT21} aims to determine whether a given data sample is used to train a target model.
Multiple works studied the membership inference attacks against the supervised learning~\cite{SSSS17,SZHBFB19,NSH19,LZ21,CTCP21,HWWBSZ21}.
Shokri et al.~\cite{SSSS17} proposed the first black-box membership inference attack against machine learning models by leveraging multiple shadow models and attack models.
The attack model takes a sample's posteriors generated from the target model as the input and predicts whether it is a member or not.
Salem et al.~\cite{SZHBFB19} relaxed the assumption from Shokri et al.~\cite{SSSS17} and proposed novel model-independent and dataset-independent membership inference attacks.
Nasr et al.~\cite{NSH19} studied the white-box membership inference attacks in both centralized and federated learning settings.
Li and Zhang~\cite{LZ21} and Choo et al.~\cite{CTCP21} concentrated on a more restricted attack scenario (called label-only attack) where the target model only returns the predicted labels instead of posteriors when the adversary queries the target model with given samples.
Roughly speaking, their proposed label-only attacks aim to infer a given sample's membership status via comparing a pre-defined threshold with the scale of adversarial perturbation that needs to be added to the given sample to change the target model's predicted label.
However, these membership inference attacks are tailored to supervised learning and we show that semi-supervised learning is more vulnerable to our proposed data augmentation-based membership inference attack compared with existing membership inference attacks.

% ----------------------------------------------------
\section{Conventional Membership Inference Attacks}
\label{section:mia}
% ----------------------------------------------------

In membership inference attack, the adversary aims to determine whether a given data sample $x$ belongs to the target model $\TargetModel$'s training dataset or not given the adversary's background knowledge $\Knowledge$.
A data sample $x$ is called \emph{member} (or \emph{non-member}) if it belongs to (or does not belong to) the training dataset of the target model $\TargetModel$.
Formally, we define the membership inference attack as $\AttackModel: x,\TargetModel, \Knowledge \rightarrow \{0, 1\}$, where the attack $\AttackModel$ is essentially a mapping function and 1 (or 0) means the data sample $x$ is a member (or non-member).

% ----------------------------------------------------
\subsection{Threat Model}
\label{subsection:threat_model}
% ----------------------------------------------------

Given a target model $\TargetModel$, we first assume that the adversary only has black-box access to it, which means that the adversary can only query the target model with a data sample $x$ and obtain the target model's prediction on it (denoted as posteriors).
Note that in this paper we consider the black-box attack since it is the most difficult and practical real-world scenario.

Following previous work~\cite{SSSS17,HZ21,SM21}, we assume that the adversary has a \emph{shadow dataset} $\ShadowDataset$ that has the same distribution as the target model $\TargetModel$'s training dataset $\TargetTrain$.
The adversary can use the shadow dataset $\ShadowDataset$ to train a \emph{shadow model} $\ShadowModel$, which mimics the behavior of the target model $\TargetModel$ to better conduct the attacks.
Also, we assume that the shadow model $\ShadowModel$ has the same architecture as the target model.
Such assumption is realistic as: (1) The adversary can leverage the same machine learning service to train the shadow model and (2) The adversary can perform hyperparameter stealing attacks~\cite{OASF18,WG18} to obtain the target model's architecture.

% ----------------------------------------------------
\subsection{Methodology}
\label{subsection:methodology}
% ----------------------------------------------------

Generally speaking, the membership inference attack pipeline usually consists of three major components, i.e., shadow training, constructing attack training dataset, and attack model training or performing the attack.

\mypara{Shadow Training} 
Shadow training~\cite{SSSS17,NSH18,SZHBFB19} aims to train shadow models to mimic the behavior of the target model based on the adversary's background knowledge.
Specifically, the adversary first evenly splits the shadow dataset $\ShadowDataset$ into two disjoint parts, i.e., shadow training data $\ShadowTrain$ and shadow testing data $\ShadowTest$.
The adversary then uses the $\ShadowTrain$ to train a shadow model $\ShadowModel$ that mimics the behavior of the target model $\TargetModel$.

\mypara{Constructing Attack Training Dataset}
To construct the training dataset for the attack model, the adversary first  uses $\ShadowTrain$ (contains members) and $\ShadowTest$ (contains non-members) to query the shadow model $\ShadowModel$ and obtain the corresponding posteriors.
Following Salem et al.~\cite{SZHBFB19}, we leverage the descendingly sorted posteriors as the inputting features for the attack model.
Finally, we assign the membership status 1/0 for members/non-members as labels.

\mypara{Training Neural Network-based Attack Model}
For neural network-based attacks~\cite{SSSS17,SZHBFB19} (denoted as $\AttackModel_{NN}$), the adversary aims to train a neural network-based attack model to distinguish members and non-members given the posteriors generated by the target model $\TargetModel$.
After constructing the attack training dataset, the adversary trains an NN-based attack model on the constructed training dataset.
Following previous works~\cite{SSSS17,SZHBFB19,HZ21,LJQG21}, we consider a multi-layer perceptron (MLP) as the neural network architecture for the attack model.
Once the attack model is trained, it can be used by the adversary to predict whether a given data sample $x$ is a member or non-member.

\mypara{Metric-based Attacks}
Metric-based attacks~\cite{YGFJ18,SSM19,LF20,SM21} also require the adversary to train a shadow model $\ShadowModel$.
Unlike NN-based attacks that require training an attack model, metric-based attacks design a specific metric and calculate a threshold over the metrics by querying the shadow model $\ShadowModel$ with $\ShadowTrain$ and $\ShadowTest$.
We adopt four state-of-the-art metric-based attacks following Song and Mittal.~\cite{SM21}: (1) Prediction correctness ($\AttackModel_{Corr}$) which considers a sample as a member if the label is correctly predicted by the target model; (2) Prediction confidence ($\AttackModel_{Conf}$) which judges a sample as a member if the prediction probability at the ground truth class is larger than a pre-defined threshold (learned from the shadow model); 
(3) prediction entropy ($\AttackModel_{Ent}$) which considers a sample as a member if the entropy of the prediction is smaller than a pre-defined threshold (learned from the shadow model); and (4) Modified prediction entropy ($\AttackModel_{Ment}$) which is similar to (3) but modifies the entropy function and combines the ground truth label as a new metric.

% ----------------------------------------------------
\section{Our Method}
\label{section:our_method}
% ----------------------------------------------------

The main difference between SSL methods and supervised learning methods is that SSL methods leverage a large amount of unlabeled samples together with a small amount of labeled samples to train the model. 
Recall that state-of-the-art SSL methods~\cite{XDDHN17,SBCZZRCKL20,ZWHWWOS21} leverage both weak and strong data augmentations to the unlabeled samples during the training. 
The key idea of these SSL methods is to train the model that maximizes the model's prediction agreement on weakly and strongly augmented views that come from the same unlabeled sample. 
In other words, for an unlabeled training sample, the trained model may tend to output more similar posteriors for its weakly and strongly augmented views.
While for labeled training samples, the trained model may output similar posteriors for different weakly augmented views from the same sample since those posteriors result in the same predicted label. 
This observation may also hold for unlabeled samples since the posteriors of the same training unlabeled sample tend to produce the same ``pseudo label''.
Intuitively speaking, the target model $\TargetModel$ may output similar (or dissimilar) posteriors for different weakly and/or strongly augmented views of member (or non-member).

Based on the above intuition, we propose a data augmentation-based membership inference attack (denoted as $\AttackModel_{DA}$) tailored to state-of-the-art SSL methods.
$\AttackModel_{DA}$ follows the similar pipeline as NN-based attack $\AttackModel_{NN}$, i.e., shadow training and training an NN-based attack model.

\begin{figure}[t]
\centering
\includegraphics[width=\columnwidth]{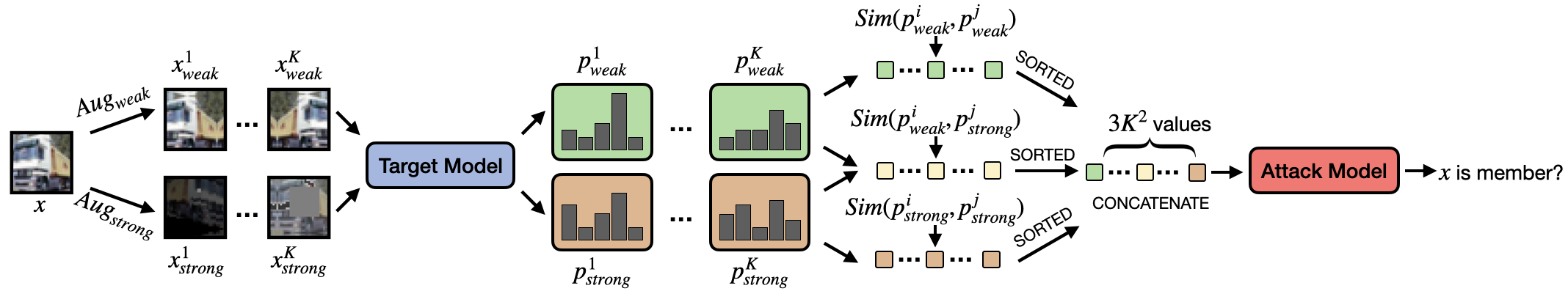}
\caption{Overview of our data augmentation based attack $\AttackModel_{DA}$.}
\label{figure:attack_pipeline}
\end{figure}

However, our attack $\AttackModel_{DA}$ extracts membership features (i.e. the input for the attack model) in a different way from the attack $\AttackModel_{NN}$. Specifically, given a data sample $x$, we first generate $K$ weakly augmented and $K$ strongly augmented views of it, respectively. 
Then we use the augmented views to query the shadow model to obtain output posteriors. 
After that, we calculate three similarity matrices among: (1) $K$ posteriors of weakly augmented views themselves, (2) $K$ posteriors of strongly augmented views themselves, and (3) $K$ posteriors of weakly augmented views and $K$ posteriors of strongly augmented views, based on a predefined similarity metrics (e.g., JS Distance, Cosine Distance, etc.).
Then we obtain three similarity matrices where each of them contains $K^2$ similarity values.
We expand each similarity matrix into a vector and sort the values in each vector in descending order, respectively.
Then we concatenate them together, and finally obtain a vector with $3K^2$ values.
The obtained vectors are then assigned with the membership status as the labels.
Once the attack model is trained, to determine whether a sample belongs to the target model's training dataset, we again generate $K$ weakly and $K$ strongly augmented views of it to query the target model, generate the attack input to query the attack model, and obtain its membership prediction. 
\autoref{figure:attack_pipeline} shows the overview of $\AttackModel_{DA}$ and the detailed algorithm is shown in Algorithm \autoref{algorithml1} in the supplemental material.

% ----------------------------------------------------
\section{Evaluation}
\label{section:evaluation}
% ----------------------------------------------------

% ----------------------------------------------------
\subsection{Experimental Setup}
% ----------------------------------------------------

\mypara{Dataset Configuration}
We evaluate the performance of target models and membership inference attacks on three commonly used SSL datasets, i.e., SVHN, CIFAR10, and CIFAR100.
For each dataset, we first randomly split it into four equal parts, i.e., $\TargetTrain$, $\TargetTest$, $\ShadowTrain$, and $\ShadowTest$.
We leverage $\TargetTrain$ to train the target model and consider the samples from $\TargetTrain$ as the members of the target model.
Samples in $\TargetTest$ are considered as the non-members of the target model.
$\ShadowTrain$ is used to build up the shadow model.
Both $\ShadowTrain$ and $\ShadowTest$ are used to train the attack model.
Note that the $\TargetTrain$ is smaller than the original training dataset (e.g., for CIFAR10, $\TargetTrain$ contains 15,000 samples while the original training dataset contains 50,000 samples), which may lead to lower target model performance.

\mypara{Metric}
We follow previous work~\cite{SBCZZRCKL20,XDHLL20,ZWHWWOS21,SSSS17,HZ21} and adopt testing accuracy as the evaluation metric for target model performance.
Regarding the attack, we leverage AUC as the evaluation metric~\cite{WGCS21,LZ21} as we aim to quantify both the general membership privacy risk for members vs. non-members and the separate privacy risks for labeled/unlabeled members vs. non-members (unbalanced).

\mypara{Target Model}
For a fair comparison, we apply the same hyperparameters for FixMatch, UDA, and FlexMatch.
Specifically, we apply SGD optimizer.
The initial learning rate is set to 0.03 with a cosine learning rate decay which sets the learning rate to $\eta \;cos(\frac{\pi k}{2N})$, where $\eta$ is the initial learning rate, $k$ is the current training step, and $N$ is the total number of training steps.
We set $N=100 \times 2^{10}$.
We leverage an exponential moving average of model parameters with the momentum of 0.999.
The labeled batch size (i.e., the batch size of the labeled data) is set to 64 and the ratio of unlabeled batch size to the labeled batch size is set to 7.
Note that the threshold $\tau$ is set to 0.8 for UDA and 0.95 for FixMatch and FlexMatch following the original papers.
We apply RandAugment~\cite{CZSL20} as the strong augmentation method in our experiments (see \autoref{appendix:augmentation} in the supplementary material).
Regarding the model architectures, we leverage Wide ResNet (WRN)~\cite{ZK16} with a widen factor of 2 as the target model architecture and we also investigate different widen factors in our ablation studies (see \autoref{subsection:ablation_study_target_model}).

\mypara{Attack Model}
We apply a 3-layer MLP with 64, 32, and 2 hidden neurons for each layer as the attack model's architecture.
We train the attack model for 100 epochs using Adam optimizer with the learning rate of 0.001 and the batch size of 256.
For our proposed attack, we set the number of augmented views used to query the target model to 10 and leverage JS Distance as the similarity function.
Note that we also evaluate different numbers of augmented views and different similarity functions in our ablation studies (see \autoref{subsection:ablation_study_attack_model}).

% ----------------------------------------------------
\subsection{Target Model Performance}
% ----------------------------------------------------

\begin{figure}[!t]
\centering
\begin{subfigure}{0.30\columnwidth}
\includegraphics[width=\columnwidth]{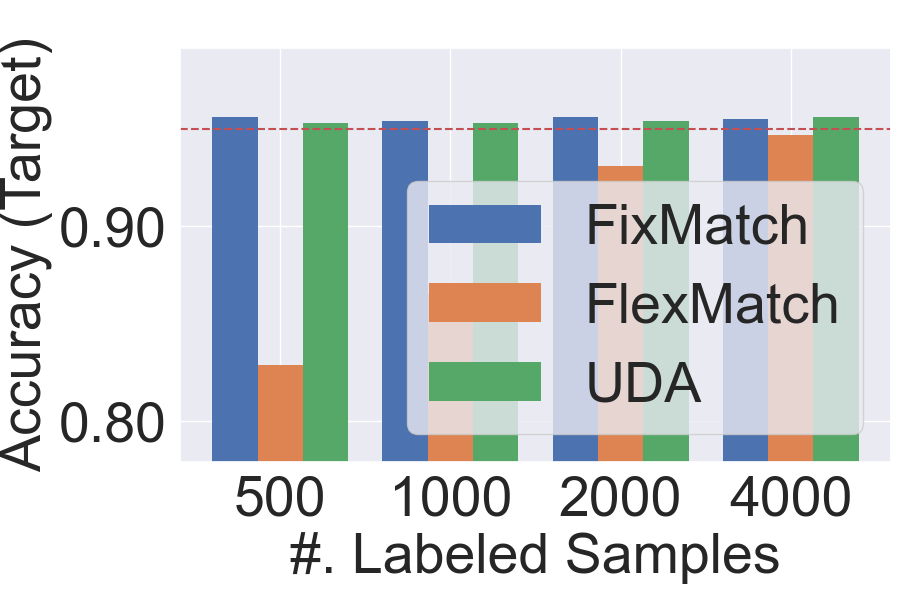}
\caption{SVHN}
\label{figure:svhn_target_test_acc}
\end{subfigure}
\begin{subfigure}{0.30\columnwidth}
\includegraphics[width=\columnwidth]{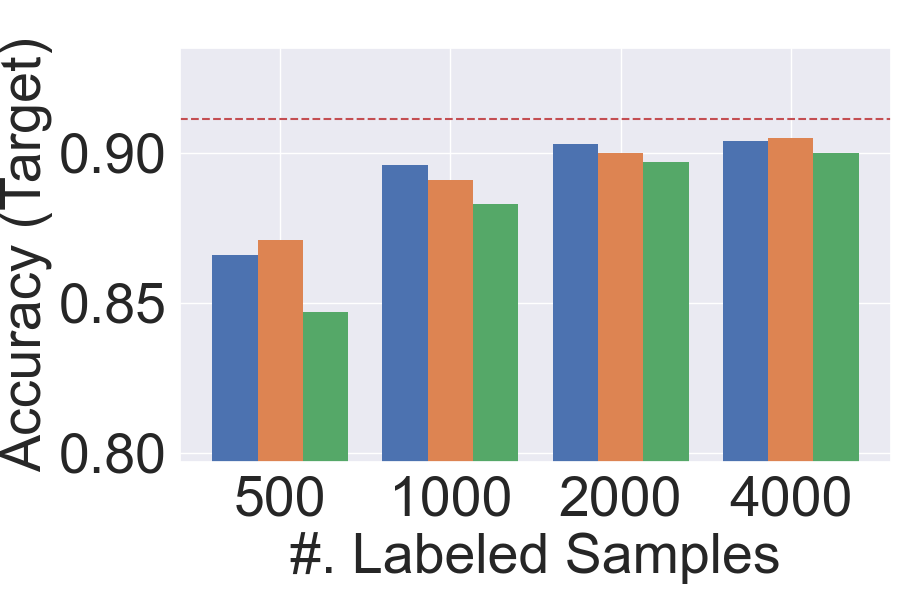}
\caption{CIFAR10}
\label{figure:cifar10_target_test_acc}
\end{subfigure}
\begin{subfigure}{0.30\columnwidth}
\includegraphics[width=\columnwidth]{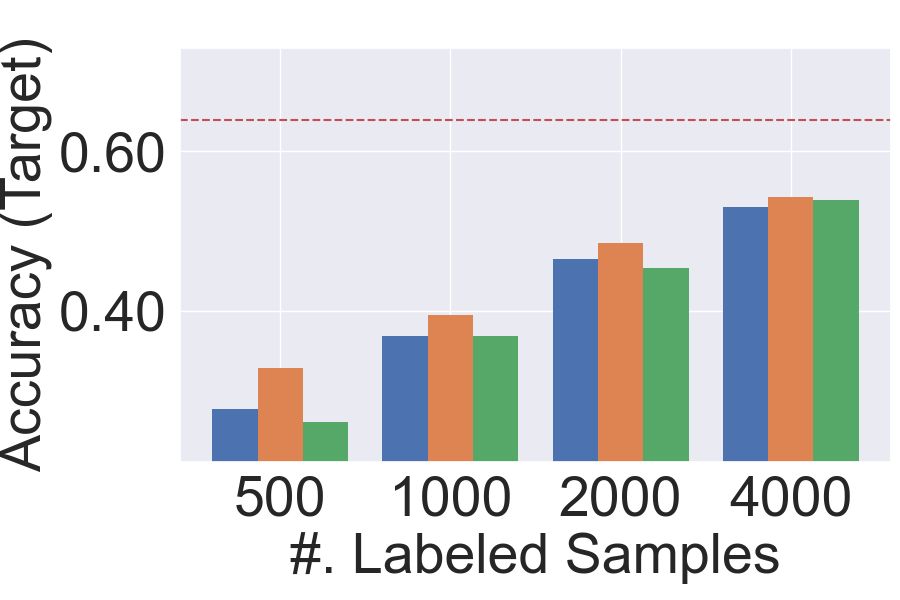}
\caption{CIFAR100}
\label{figure:cifar100_target_test_acc}
\end{subfigure}
\caption{Testing accuracy on the original classification tasks. 
Note that the red dash line denotes the performance of supervised models.}
\label{figure:target_test_acc}
\end{figure}

We first evaluate the performance of the supervised models and the SSL models on the original classification tasks using $\TargetTest$.
We use the full  $\TargetTrain$ to train the supervised models, while we use a small portion of labeled samples and treat the remaining samples as unlabeled ones in $\TargetTrain$ when training the SSL models. We observe that SSL with more labeled samples can achieve better performance on the original classification tasks. 
For instance, on \autoref{figure:cifar10_target_test_acc}, when the target model is FixMatch trained on CIFAR10, the classification accuracy is 0.866, 0.896, 0.903, and 0.904 with 500, 1,000, 2,000, and 4,000 labeled samples, respectively.
This is expected as more labeled samples help the target model to better learn the decision boundary at the early stage.
Another observation is that for a more complicated task, it may require more labeled samples to achieve comparable performance as the supervised models.
We consider SVHN, CIFAR10, and CIFAR100 have increasing difficulty levels.
Take models trained by UDA as a case study (green bar in \autoref{figure:target_test_acc}), on SVHN, with only 500 labeled samples, the testing accuracy is 0.953, which is even better than the supervised model (0.951).
We suspect the reason is that 500 labeled samples is enough to learn a relatively accurate decision boundary and the strong data augmentation used in SSL methods can better help the model to generalize to the unseen data.
On the other hand, on CIFAR10 and CIFAR100, it may require 1,000 and 4,000 labeled samples to catch up with the performance of the supervised model.
Such observation indicates that a larger portion of labeled data is still helpful for a more complicated task.

% ----------------------------------------------------
\subsection{Membership Inference Attack Performance}
% ----------------------------------------------------

\begin{figure}[!t]
\centering
\begin{subfigure}{0.90\columnwidth}
\includegraphics[width=\columnwidth]{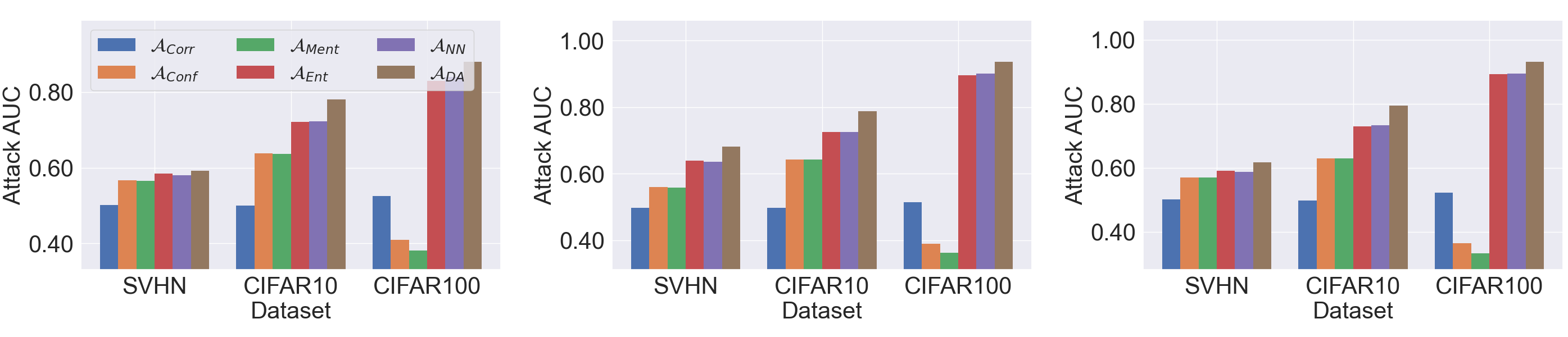}
\caption{Attack AUC}
\label{figure:combine_WideResNet_500_att_test_auc}
\end{subfigure}
\begin{subfigure}{0.90\columnwidth}
\includegraphics[width=\columnwidth]{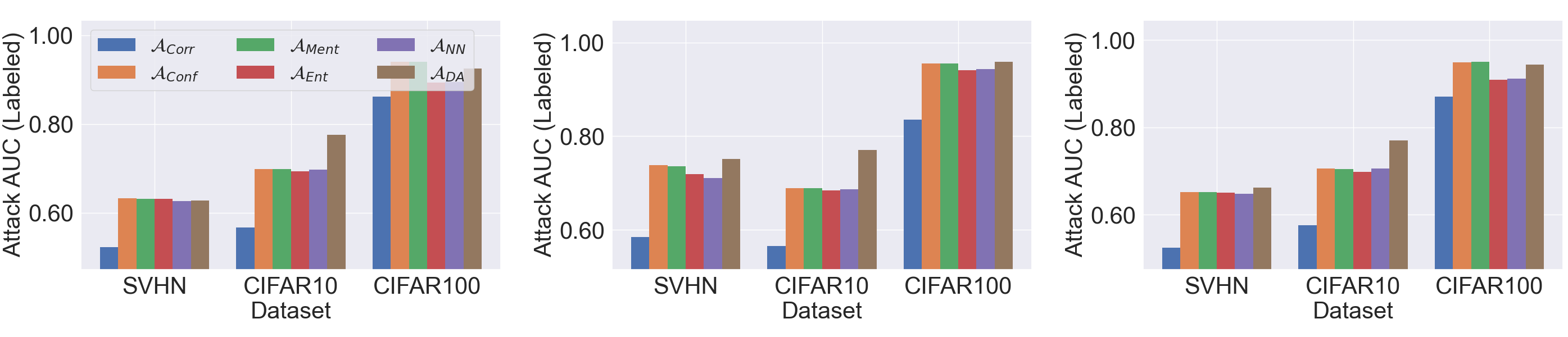}
\caption{Attack AUC (Labeled)}
\label{figure:combine_WideResNet_500_labeled_auc}
\end{subfigure}
\begin{subfigure}{0.90\columnwidth}
\includegraphics[width=\columnwidth]{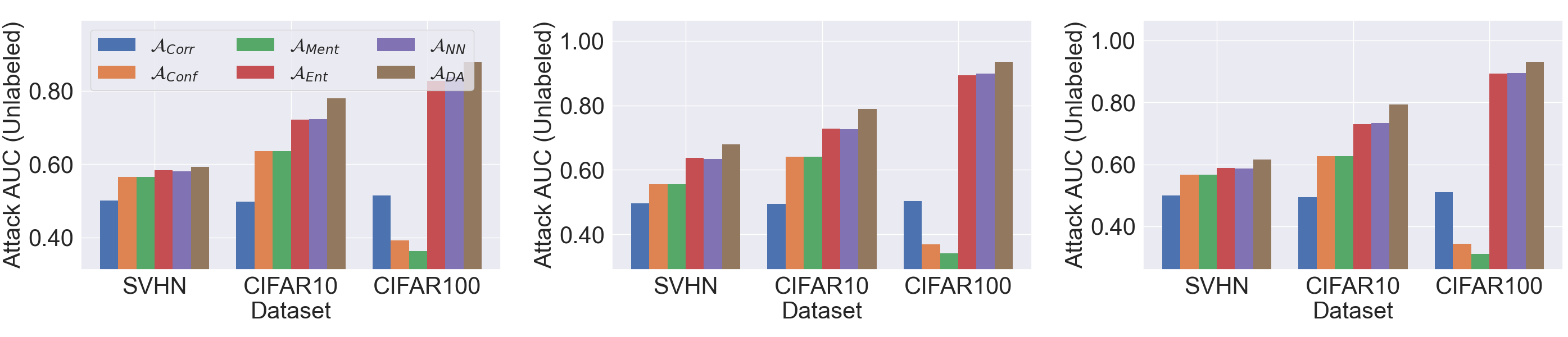}
\caption{Attack AUC (Unlabeled)}
\label{figure:combine_WideResNet_500_unlabeled_auc}
\end{subfigure}
\caption{The AUC of membership inference attacks against models trained by different SSL methods with 500 labeled samples. 
The first to third columns denote the models trained by FixMatch, FlexMatch, and UDA, respectively.}
\label{figure:attack_performance_500}
\end{figure}

We then evaluate the performance of different membership inference attacks on SSL models.
The results are summarized in \autoref{figure:attack_performance_500}.
Note that we leverage AUC as the attack evaluation metric to better quantify the privacy leakage of all training data (first row) as well as the separate privacy leakage of labeled (second row) and unlabeled (third row) training data.
We find that for the baseline attacks (i.e., except our $\AttackModel_{DA}$), $\AttackModel_{NN}$ and $\AttackModel_{Ent}$ perform the best, while other attacks like $\AttackModel_{Corr}$, $\AttackModel_{Conf}$, and $\AttackModel_{Ment}$ are less effective.
For instance, on FlexMatch trained on CIFAR10 with 500 labeled samples (the middle one of \autoref{figure:combine_WideResNet_500_att_test_auc}), the attack AUC is 0.726 for both $\AttackModel_{NN}$ and $\AttackModel_{Ent}$, while only 0.497, 0.643, and 0.642 for $\AttackModel_{Corr}$, $\AttackModel_{Conf}$, and $\AttackModel_{Ment}$.
To better investigate the reason behind this, we further measure the attack AUC for labeled data and unlabeled data, respectively.
We find that $\AttackModel_{Conf}$ and $\AttackModel_{Ment}$ achieve even better performance on labeled training samples than $\AttackModel_{NN}$ and $\AttackModel_{Ent}$.
For instance, for FlexMatch trained on CIFAR100, the AUC (labeled) for $\AttackModel_{Conf}$ and $\AttackModel_{Ment}$ are both 0.955, while only 0.944 and 0.941 for $\AttackModel_{NN}$ and $\AttackModel_{Ent}$.
This is expected as the labeled sample has a higher confidence score on its ground-truth label, which facilitates the attacks that leverage such information.
However, this is not the case for the unlabeled samples.
As we can observe that, for FlexMatch trained on CIFAR100, the AUC (unlabeled) is only 0.370 and 0.341 for $\AttackModel_{Conf}$ and $\AttackModel_{Ment}$, but 0.899 and 0.894 for $\AttackModel_{NN}$ and $\AttackModel_{Ent}$.
This indicates that, for the unlabeled samples, the model may give similar correctness predictions on both unlabeled training samples and testing samples, which makes it harder to differentiate them.
However, the model will give more confident predictions on unlabeled training samples than on testing samples, which results in better performance for $\AttackModel_{NN}$ and $\AttackModel_{Ent}$.

On the other hand, we also observe that the data augmentation-based attack $\AttackModel_{DA}$ achieves consistently better overall performance on all datasets and SSL methods than those baseline attacks.
Moreover, $\AttackModel_{DA}$ works better in determining the membership of unlabeled training samples.
For instance, on FixMatch trained on CIFAR10, the unlabeled AUC is 0.780 for $\AttackModel_{DA}$ while only 0.722 for the best baseline attack ($\AttackModel_{NN}$).
This is because $\AttackModel_{DA}$ unveils the pattern that the predictions of a sample's weak and strong augmented views should be closer if the sample is an unlabeled sample used during the training.

% ----------------------------------------------------
\subsection{What Determines Membership Inference Attack in SSL}
\label{subsection:epoch_wised_performance}
% ----------------------------------------------------

The effectiveness of membership inference attacks has been largely credited to the intrinsic overfitting phenomenon of the ML model~\cite{SSSS17,SZHBFB19}.
Here overfitting denotes the model's training accuracy minus its testing accuracy.
Such assumption has been verified on various ML models~\cite{SSSS17,NSH18,SZHBFB19,HZ21}.
However, it is unclear whether such assumption still holds for SSL.
If not, what is the reason for models trained by SSL to be vulnerable to membership inference attacks?

From \autoref{figure:attack_performance_500}, we find that $\AttackModel_{Ent}$ achieves good performance in predicting the membership status of a sample, which gives us the hint that the members' and non-members' predictions may have different entropy distributions.
Here we leverage the JS Distance to quantify the difference between the entropy distribution of members' and non-members' predictions (we denote this measure as JS Distance (Entropy)).

\begin{figure*}[!t]
\centering
\begin{subfigure}{0.30\columnwidth}
\includegraphics[width=\columnwidth]{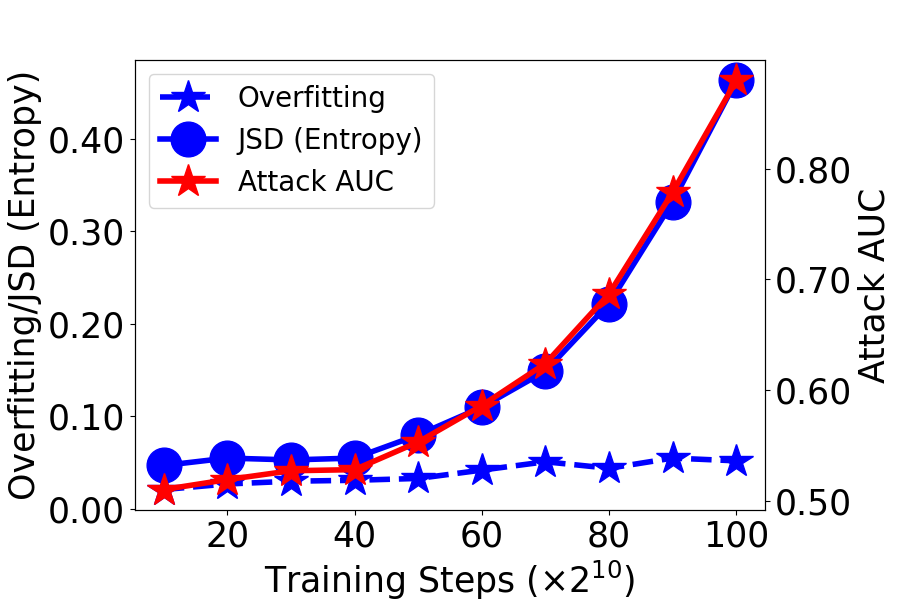}
\caption{FixMatch}
\label{figure:fixmatch_cifar100_WideResNet_500_epoch_performance}
\end{subfigure}
\begin{subfigure}{0.30\columnwidth}
\includegraphics[width=\columnwidth]{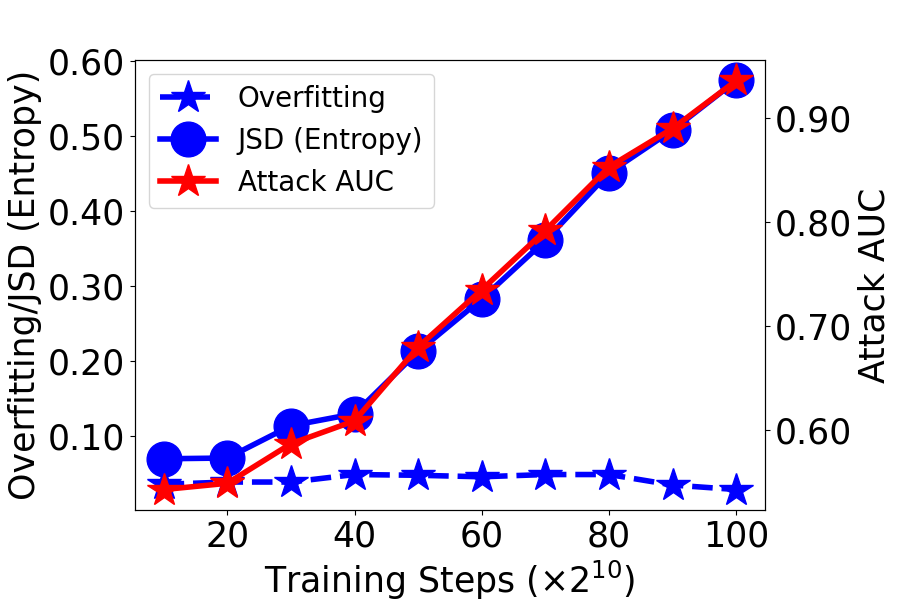}
\caption{FlexMatch}
\label{figure:flexmatch_cifar100_WideResNet_500_epoch_performance}
\end{subfigure}
\begin{subfigure}{0.30\columnwidth}
\includegraphics[width=\columnwidth]{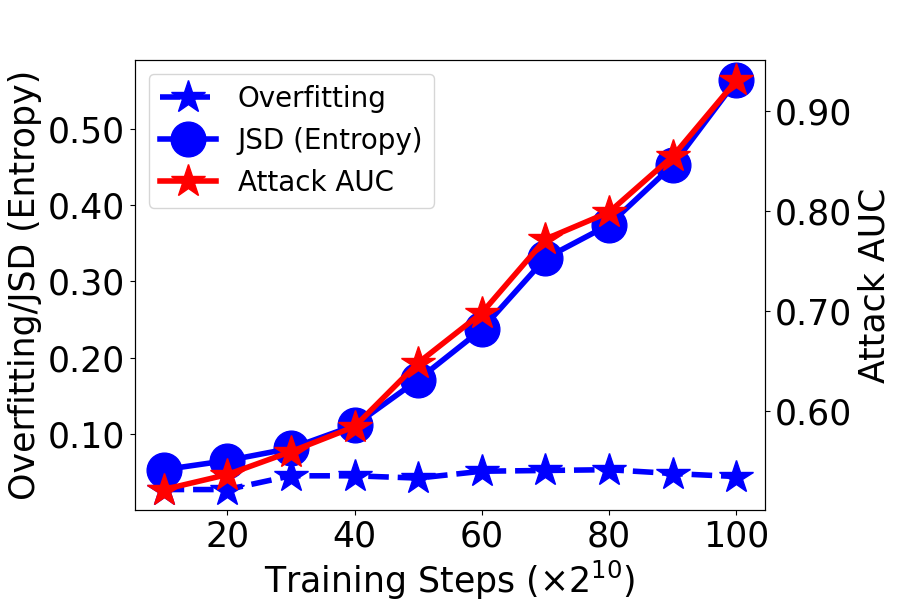}
\caption{UDA}
\label{figure:uda_cifar100_WideResNet_500_epoch_performance}
\end{subfigure}
\caption{The overfitting/JS Distance (Entropy) and attack AUC with respect to different training steps. 
The target model is trained on CIFAR100 with 500 labeled samples. 
Note that we consider the attack AUC of $\AttackModel_{DA}$, which is the strongest attack.}
\label{figure:cifar100_epoch_performance_500}
\end{figure*}

To better quantify the correlation between different factors (e.g., overfitting, JS Distance (Entropy)) and the attack performance, we measure them under different training steps of the target models.
Note that here we consider the $\AttackModel_{DA}$ as it performs the best in membership inference.
\autoref{figure:cifar100_epoch_performance_500} shows the results of models trained by different SSL methods on the CIFAR100 with 500 labeled samples.
The results for models trained on different datasets and with different numbers of labeled samples are shown in \autoref{appendix:epoch_performance} in supplementary materials.

In \autoref{figure:cifar100_epoch_performance_500}, we observe that during the whole training procedure, the models trained by SSL have nearly 0 overfitting, which means that the models can always generalize well to the unseen data.
However, we find that the attack AUC keeps increasing during the training.
This indicates that the success of membership inference attacks is not necessarily related to the high overfitting level, which is overlooked by previous research. 
On the other hand, we observe that the JS Distance (Entropy) does increase during the training, which means that although the model does not predict more accurately to the member samples (mainly unlabeled samples) than the non-member samples, the model indeed makes a more confident prediction on member samples (i.e., with lower entropy of prediction).
Our observation reveals that the models trained by SSL indeed ``memorize'' the training data.
However, such memorization does not reflect in the overfitting, i.e., the gap between training and testing accuracy.
Instead, it reflects in the more confident prediction of the members than the non-members.

% ----------------------------------------------------
\subsection{Ablation Study (Attack Model)}
\label{subsection:ablation_study_attack_model}
% ----------------------------------------------------

\begin{figure}[!t]
\centering
\begin{subfigure}{0.30\columnwidth}
\includegraphics[width=\columnwidth]{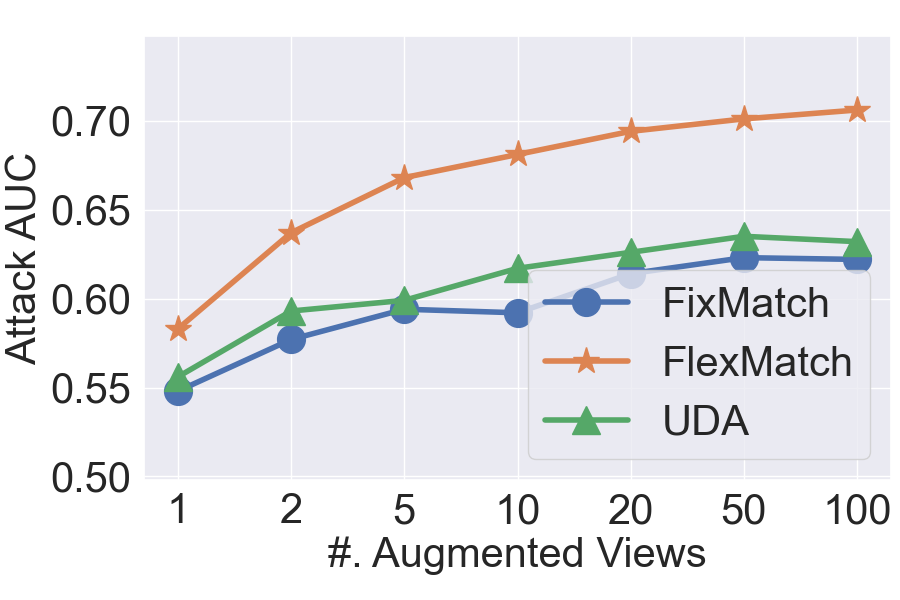}
\caption{SVHN}
\label{figure:svhn_500_ablation_view_test_auc}
\end{subfigure}
\begin{subfigure}{0.30\columnwidth}
\includegraphics[width=\columnwidth]{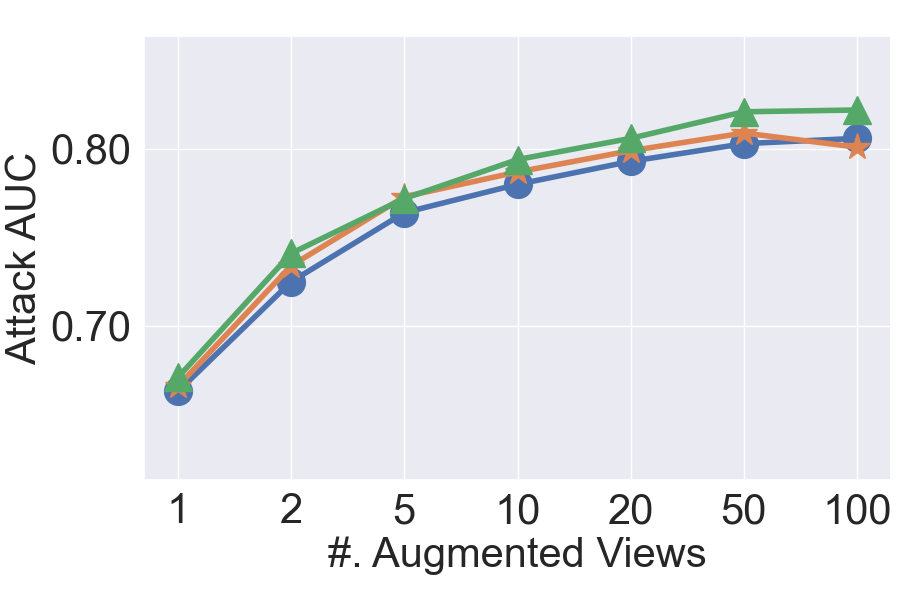}
\caption{CIFAR10}
\label{figure:cifar10_500_ablation_view_test_auc}
\end{subfigure}
\begin{subfigure}{0.30\columnwidth}
\includegraphics[width=\columnwidth]{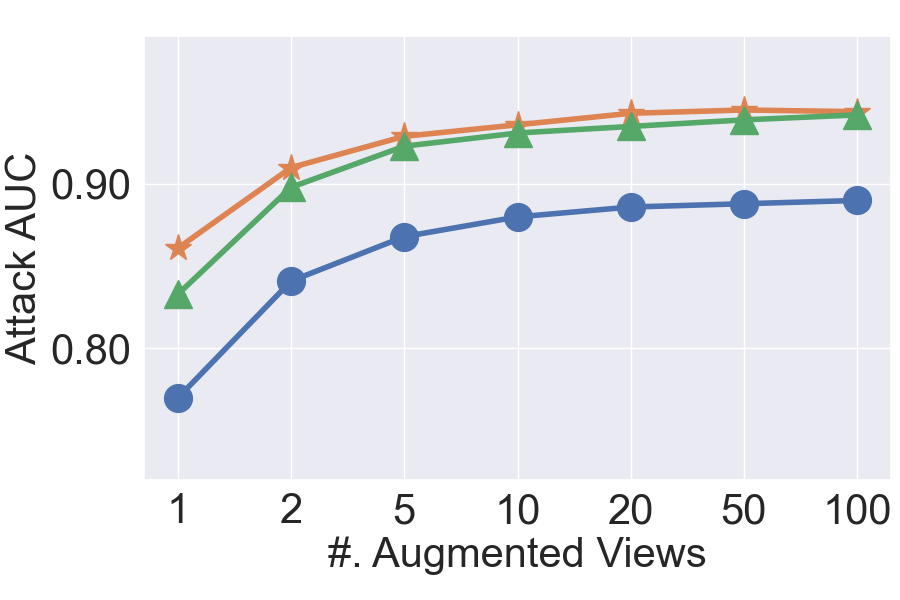}
\caption{CIFAR100}
\label{figure:cifar100_500_ablation_view_test_auc}
\end{subfigure}
\caption{The attack AUC of $\AttackModel_{DA}$  with different numbers of augmented views to query the target model. 
The target model is trained with 500 labeled samples.}
\label{figure:500_ablation_view_test_auc}
\end{figure}

\mypara{Number of Views}
We first investigate how the attack performance would be affected by different numbers of views generated by the weak and strong augmentations to query the target model.
To this end, for the SSL methods trained on different datasets with only 500 labeled samples, we range the number of views from 1 to 100 and the attack performance is shown in \autoref{figure:500_ablation_view_test_auc}.
Note that we also show the results with 1,000, 2,000, and 4,000 labeled samples in \autoref{appendix:ablation_number_of_view} in the supplementary material.
A clear trend is that more views lead to better attack performance.
For instance, for FixMatch trained on CIFAR10 with 500 labeled samples (\autoref{figure:cifar10_500_ablation_view_test_auc}), the attack AUC is 0.780 with 10 augmented views, while 0.806 for 100 augmented views. 
However, we find that the attack performance increases rapidly when the number of augmented views increases from 1 to 10, but plateaus from 10 to 100.
Moreover, more views lead to more queries to the target model and higher computational cost.
We consider 10 as a suitable number of views since it achieves comparable performance to 100 while spending less query budget.

\begin{figure}[!t]
\centering
\begin{subfigure}{0.30\columnwidth}
\includegraphics[width=\columnwidth]{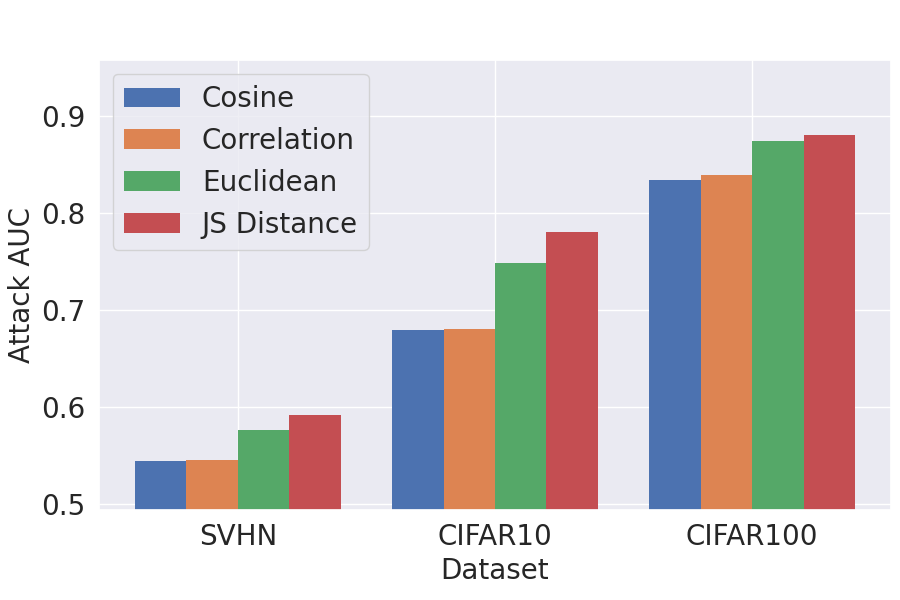}
\caption{FixMatch}
\label{figure:fixmatch_500_ablation_similarity_func_test_auc}
\end{subfigure}
\begin{subfigure}{0.30\columnwidth}
\includegraphics[width=\columnwidth]{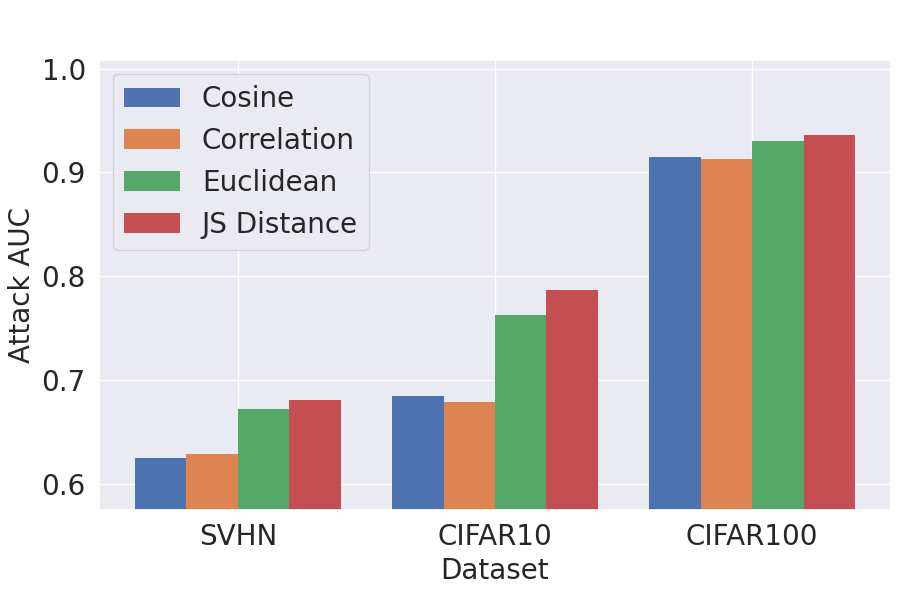}
\caption{FlexMatch}
\label{figure:flexmatch_500_ablation_similarity_func_test_auc}
\end{subfigure}
\begin{subfigure}{0.30\columnwidth}
\includegraphics[width=\columnwidth]{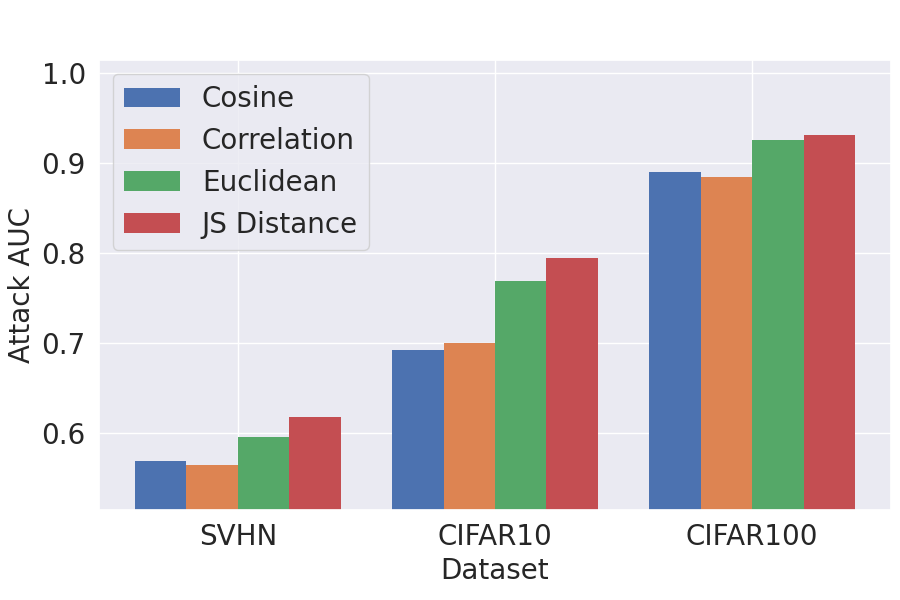}
\caption{UDA}
\label{figure:uda_500_ablation_similarity_func_test_auc}
\end{subfigure}
\caption{The attack AUC of $\AttackModel_{DA}$  with different similarity functions. 
The target model is trained with 500 labeled samples.}
\label{figure:ablation_similarity_func_test_auc_500}
\end{figure}

\mypara{Similarity Function}
Note that in our attack $\AttackModel_{DA}$, we can apply different similarity functions to measure the distance between the posteriors generated from different augmented views.
Here we evaluate 4 distance metrics, i.e., Cosine Distance, Correlation Distance, Euclidean Distance, and JS Distance.
The result for FixMatch, FlexMatch, and UDA trained on three different datasets with 500 labeled samples are summarized in \autoref{figure:ablation_similarity_func_test_auc_500}.
Note that we also show the results with 1,000, 2,000, and 4,000 labeled samples in \autoref{appendix:ablation_similarity_function} in the supplementary material.
We find that JS Distance consistently outperforms the other three distance metrics and achieves the best performance.
For instance, FixMatch trained on CIFAR10, the attack AUC is 0.679, 0.682, 0.749, and 0.780 for Cosine Distance, Correlation Distance, Euclidean Distance, and JS Distance.
We suspect the reason is that JS Distance is designed to calculate the difference between two probabilities' distributions, which may better fit our scenario as the prediction posteriors are probability as well.

Moreover, we also find that the magnitude of data augmentation and the shadow model architecture only have limited impact on the attack performance (see \autoref{appendix:ablation_attack_model} in the supplementary material for more details).

% ----------------------------------------------------
\subsection{Ablation Study (Target Model)}
\label{subsection:ablation_study_target_model}
% ----------------------------------------------------

We also investigate whether the target model's capacity and the unlabeled ratio (i.e., $\frac{batchsize (unlabeled)}{batchsize (labeled)}$ during each training step) would affect the performance.
Note that here we select FixMatch trained on CIFAR100 with 500 labeled data as a case study, since the target model's capacity and the unlabeled ratio are general to different SSL methods, and CIFAR100 with 500 labeled data is the most challenging setting to train the target model (see \autoref{figure:target_test_acc}).
We consider an adaptive adversary~\cite{JSBZG19} who is aware of the training details of the target model and can train the shadow model in the same way.

\mypara{Model Capacity}
The target model architecture we leverage in our paper is WRN28-2.
To better quantify the impact of model capacity on the target and attack performance, we vary the width of WRN28 from 1 to 8 and the results are shown in \autoref{table:ablation_model_capacity}.
We can observe that a larger model capacity, in general, leads to a better target model's performance on the original classification task, but also increases the membership risk (especially for unlabeled samples).
For instance, when the model capacity increase from WRN28-1 to WRN-28-8, the target testing accuracy increases from 0.217 to 0.305, while the attack AUC increases from 0.726 to 0.927.
One reason is that, with larger model capacity, the model can ``memorize'' more different views of data samples, which not only facilitate target tasks, but also raise the membership risk.

\begin{table}[!t]
\caption{The target model performance and attack performance ($\AttackModel_{DA}$) when the target model has different capacities. 
The target model is trained by FixMatch on CIFAR100 with 500 labeled samples.
($\star$) denotes the default setting.}
\label{table:ablation_model_capacity}
\centering
\resizebox{0.99\linewidth}{!}{
\begin{tabular}{l | c | c | c | c}
\toprule
Architecture & Test ACC & Attack AUC & Attack AUC (Labeled) & Attack AUC (Unlabeled) \\
\midrule
WRN28-1 & 0.217 & 0.726 & 0.954 & 0.718 \\
WRN28-2 ($\star$) & 0.276 & 0.874 & 0.896 & 0.873 \\
WRN28-4 & 0.299 & 0.917 & 0.910 & 0.917 \\
WRN28-8 & 0.305 & 0.927 & 0.918 & 0.927 \\
\bottomrule
\end{tabular}
}
\end{table}

\begin{table}[!t]
\caption{The target model performance and attack performance ($\AttackModel_{DA}$) when the target model leverages different unlabeled ratios during each training step. 
The target model is trained by FixMatch on CIFAR100 with 500 labeled samples.
($\star$) denotes the default setting.}
\label{table:ablation_uratio}
\centering
\begin{tabular}{l | c |c |c |c}
\toprule
Ratio & Test Acc & Attack AUC & Attack AUC (Labeled) & Attack AUC (Unlabeled) \\
\midrule
1 & 0.210 & 0.578 & 0.965 & 0.565\\
2 & 0.263 & 0.646 & 0.942 & 0.636 \\
4 & 0.273 & 0.785 & 0.946 & 0.779\\
7 ($\star$) & 0.276 & 0.874 & 0.896 & 0.873 \\
8 & 0.269 & 0.886 & 0.924 & 0.884 \\
16 & 0.247 & 0.909 & 0.913 & 0.909 \\
\bottomrule
\end{tabular}
\end{table}

\mypara{Ratio of Unlabeled Samples in Each Training Step}
We then investigate whether the unlabeled ratio (URatio) during each training step affects the attack performance.
Concretely, we vary the unlabeled ratio from 1 to 16 while training the target model and \autoref{table:ablation_uratio} summarizes the results.
We have two findings.
First, the best target model performance reaches with the default setting (7).
Second, the membership inference risk, in particular for the unlabeled data, keeps increasing when the ratio increases.
On the other hand, the membership inference risk for labeled data slightly decreases (but still in a high level) while increasing the ratio.
Therefore, a better choice may be leveraging a relatively small unlabeled ratio to achieve good target performance while reducing the membership risk for unlabeled samples.

% ----------------------------------------------------
\section{Discussion on Defenses}
% ----------------------------------------------------

We observe that the attack performance increases sharply at the late training steps (see \autoref{figure:cifar100_epoch_performance_500}), which indicates that early stopping may be a good strategy to mitigate membership inference attacks.
We take CIFAR100 with 4,000 labeled samples as a case study and show the target/attack model performance with respect to different training steps in \autoref{figure:defense_earlystopping_cifar100_4000} (in the supplementary material).
We find that there is a trade-off between model utility and membership inference performance, i.e., it may reduce both the attack performance and the target model's utility.
We note that previous work~\cite{LJQG21,SM21} also observe such a trade-off.
Besides early stopping, we also evaluate three other defenses, i.e., top-$k$ posteriors~\cite{SSSS17}, model stacking~\cite{SZHBFB19}, and DP-SGD~\cite{ACGMMTZ16}.
Our case study (see \autoref{appendix:defense} in the supplementary material) shows that early stopping achieves the best trade-off between model utility and membership inference performance.

% ----------------------------------------------------
\section{Conclusion}
% ----------------------------------------------------

In this paper, we perform the first training data privacy quantification against models trained by SSL through the lens of membership inference attack.
Empirical evaluation shows that our proposed data augmentation-based attacks consistently outperform the baseline attacks, in particular for unlabeled training data.
Moreover, we have an interesting finding that the reason leading to membership leakage in SSL is different from the commonly believed overfitting nature of ML models trained in supervised manners.
The models trained by SSL are well generalized to the testing data (i.e., with almost 0 overfitting level).
However, our attack can still successfully break the membership privacy.
The reason is that the models trained by SSL ``memorize'' the training data by giving more confident predictions on them, regardless of the ground truth labels.
We also find that early stopping can serve as a countermeasure against the attacks, but there is a trade-off between membership privacy and model utility.

\smallskip\noindent{\textbf{Acknowledgments:}} This work is partially funded by the Helmholtz Association within the project ``Trustworthy Federated Data Analytics'' (TFDA) (funding number ZT-I-OO1 4) and National Science Foundation grant No.\ 1937786.

% ----------------------------------------------------
\newpage
\bibliographystyle{splncs04}
\bibliography{normal_generated_py3}
% ----------------------------------------------------

% ----------------------------------------------------
\clearpage
\section{Supplementary Material}
% ----------------------------------------------------

% ----------------------------------------------------
\subsection{Data Augmentation}
\label{appendix:augmentation}
% ----------------------------------------------------

Models trained by FixMatch, FlexMatch, and UDA apply both weak and strong augmentation to the unlabeled samples.
For the weak augmentation, we apply random cropping with a padding of 4 and random horizontal flipping to each sample following~\cite{KSH12,ZK16}.
For the strong augmentation, we apply random augmentation~\cite{CZSL20} to each training sample, which consists of a group of augmentation operations.
Specifically, we set $N=2$ and $M=10$ where $N$ is the number of transformations to a given sample and $M$ is the magnitude of global distortion.

\subsection{Attack Performance with Different Numbers of Labeled Samples}
\begin{figure}[!t]
\centering
\begin{subfigure}{\columnwidth}
\includegraphics[width=\columnwidth]{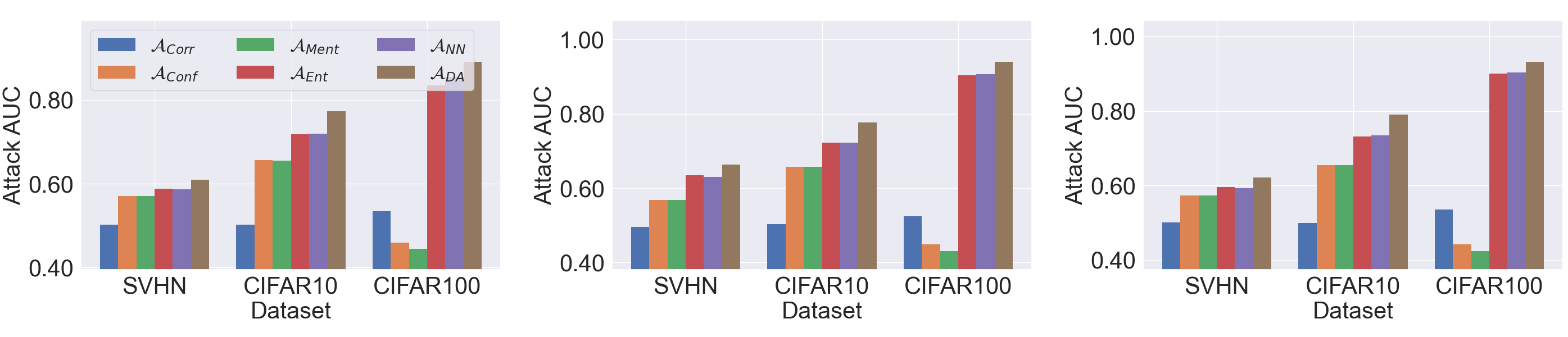}
\caption{Attack AUC}
\label{figure:combine_WideResNet_1000_att_test_auc}
\end{subfigure}
\begin{subfigure}{\columnwidth}
\includegraphics[width=\columnwidth]{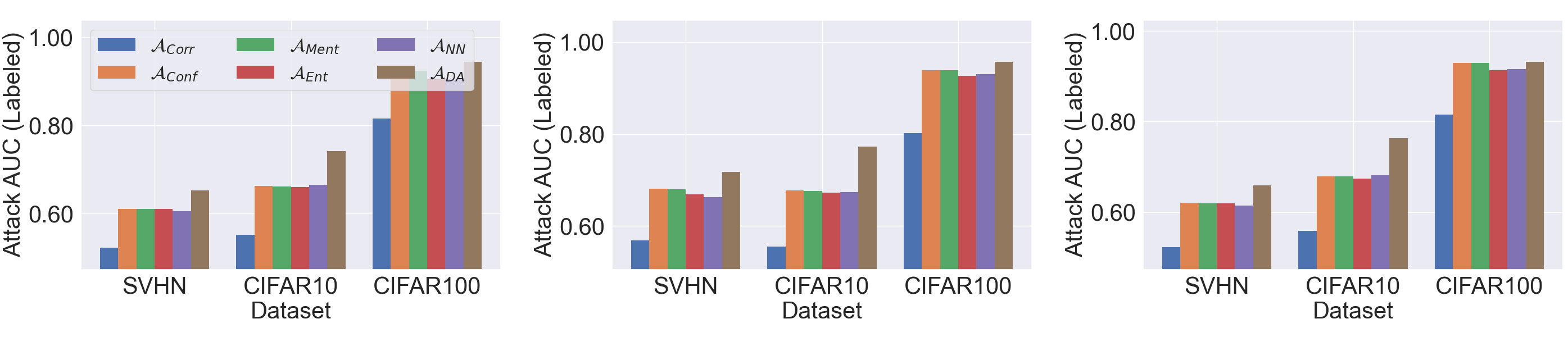}
\caption{Attack AUC (Labeled)}
\label{figure:combine_WideResNet_1000_labeled_auc}
\end{subfigure}
\begin{subfigure}{\columnwidth}
\includegraphics[width=\columnwidth]{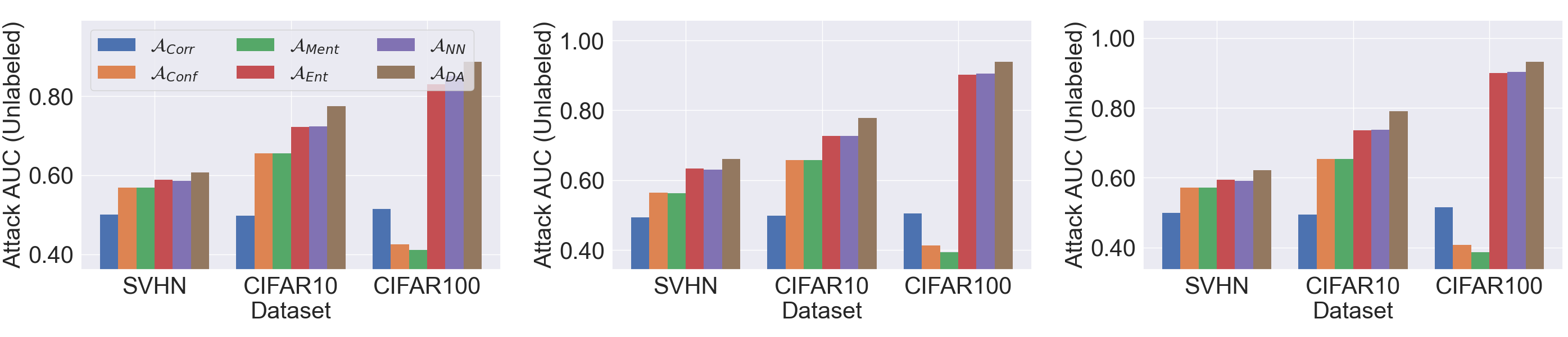}
\caption{Attack AUC (Unlabeled)}
\label{figure:combine_WideResNet_1000_unlabeled_auc}
\end{subfigure}
\caption{The AUC of membership inference attacks against models trained by different SSL methods with 1,000 label samples. 
The first to third columns denotes the model trained by FixMatch, FlexMatch, and UDA, respectively.}
\label{figure:attack_performance_1000}
\end{figure}

\begin{figure}[!t]
\centering
\begin{subfigure}{\columnwidth}
\includegraphics[width=\columnwidth]{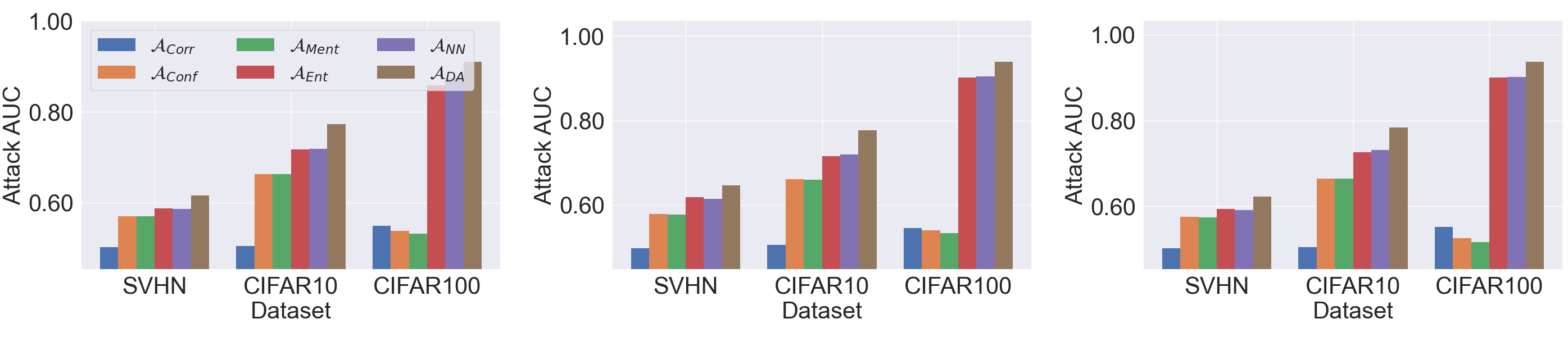}
\caption{Attack AUC}
\label{figure:combine_WideResNet_2000_att_test_auc}
\end{subfigure}
\begin{subfigure}{\columnwidth}
\includegraphics[width=\columnwidth]{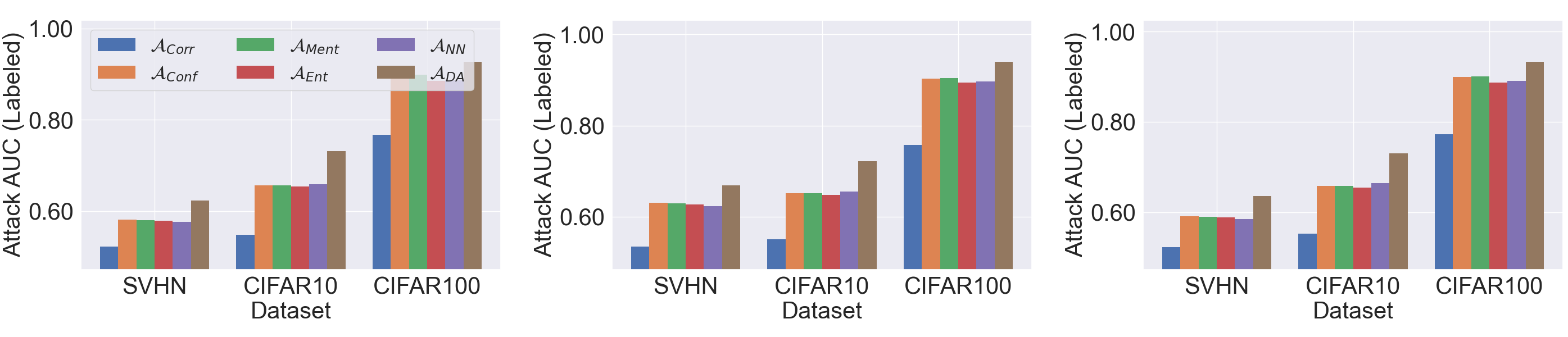}
\caption{Attack AUC (Labeled)}
\label{figure:combine_WideResNet_2000_labeled_auc}
\end{subfigure}
\begin{subfigure}{\columnwidth}
\includegraphics[width=\columnwidth]{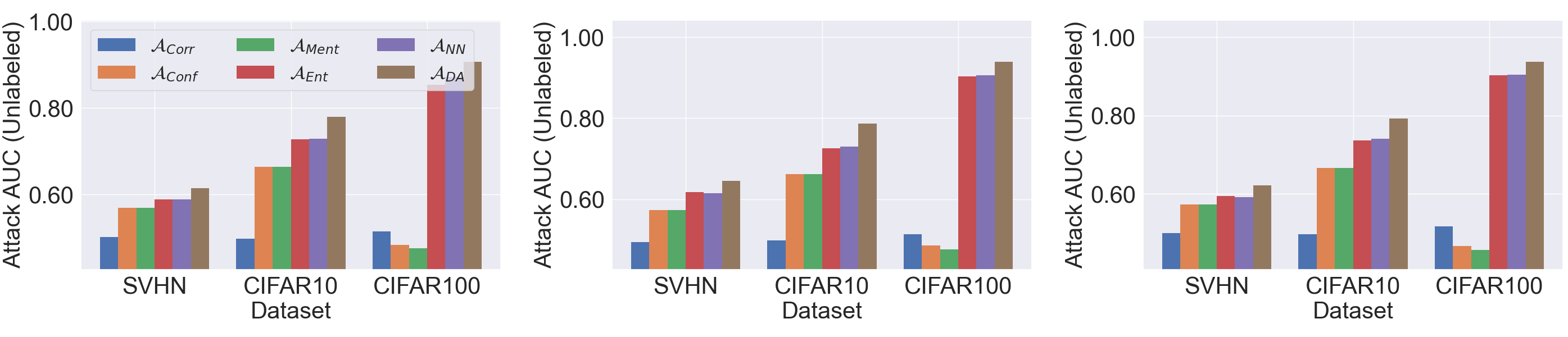}
\caption{Attack AUC (Unlabeled)}
\label{figure:combine_WideResNet_2000_unlabeled_auc}
\end{subfigure}
\caption{The AUC of membership inference attacks against models trained by different SSL methods with 2,000 label samples. 
The first to third columns denotes the model trained by FixMatch, FlexMatch, and UDA, respectively.}
\label{figure:attack_performance_2000}
\end{figure}

\begin{figure}[!t]
\centering
\begin{subfigure}{\columnwidth}
\includegraphics[width=\columnwidth]{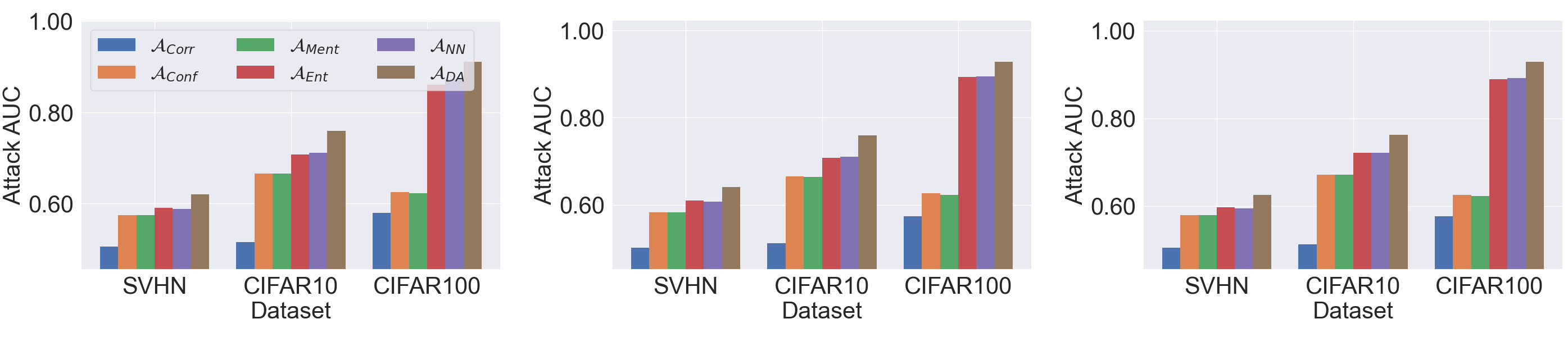}
\caption{Attack AUC}
\label{figure:combine_WideResNet_4000_att_test_auc}
\end{subfigure}
\begin{subfigure}{\columnwidth}
\includegraphics[width=\columnwidth]{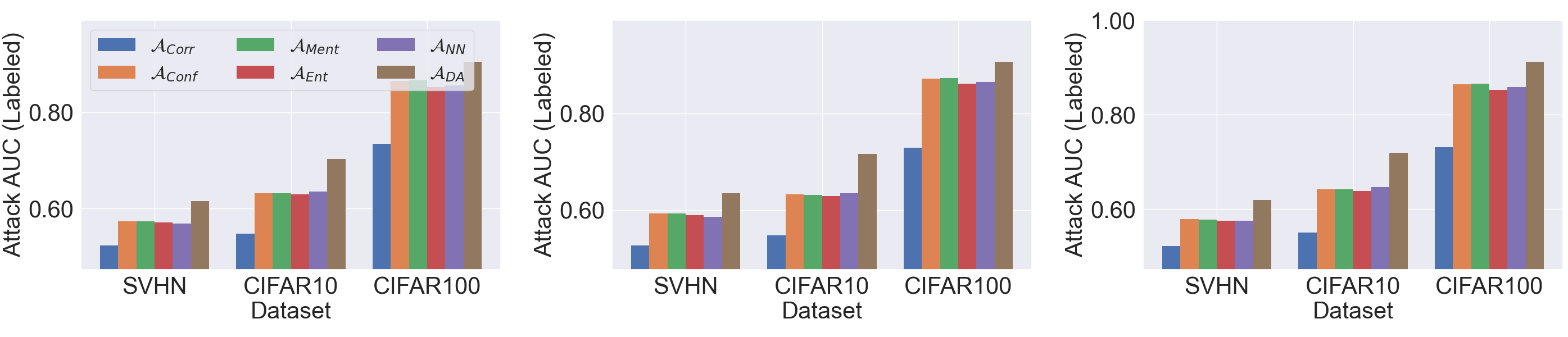}
\caption{Attack AUC (Labeled)}
\label{figure:combine_WideResNet_4000_labeled_auc}
\end{subfigure}
\begin{subfigure}{\columnwidth}
\includegraphics[width=\columnwidth]{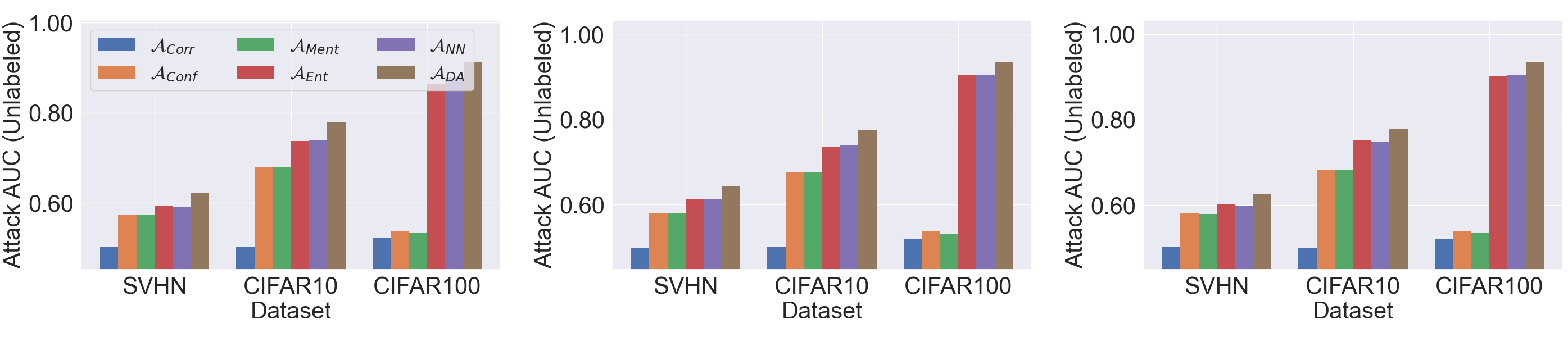}
\caption{Attack AUC (Unlabeled)}
\label{figure:combine_WideResNet_4000_unlabeled_auc}
\end{subfigure}
\caption{The AUC of membership inference attacks against models trained by different SSL methods with 4,000 label samples. 
The first to third columns denotes the model trained by FixMatch, FlexMatch, and UDA, respectively.}
\label{figure:attack_performance_4000}
\end{figure}

\autoref{figure:attack_performance_1000}, \autoref{figure:attack_performance_2000}, and \autoref{figure:attack_performance_4000} show the attack performance with 1,000, 2,000, and 4,000 labeled training data.
We observe that our proposed data augmentation-based attack $\AttackModel_{DA}$ still consistently outperforms baseline attacks.

% ----------------------------------------------------
\subsection{What Determines Membership Inference Attack in SSL with Different Numbers of Labeled Samples.}
\label{appendix:epoch_performance}
% ----------------------------------------------------

\begin{figure}[!t]
\centering
\begin{subfigure}{\columnwidth}
\includegraphics[width=\columnwidth]{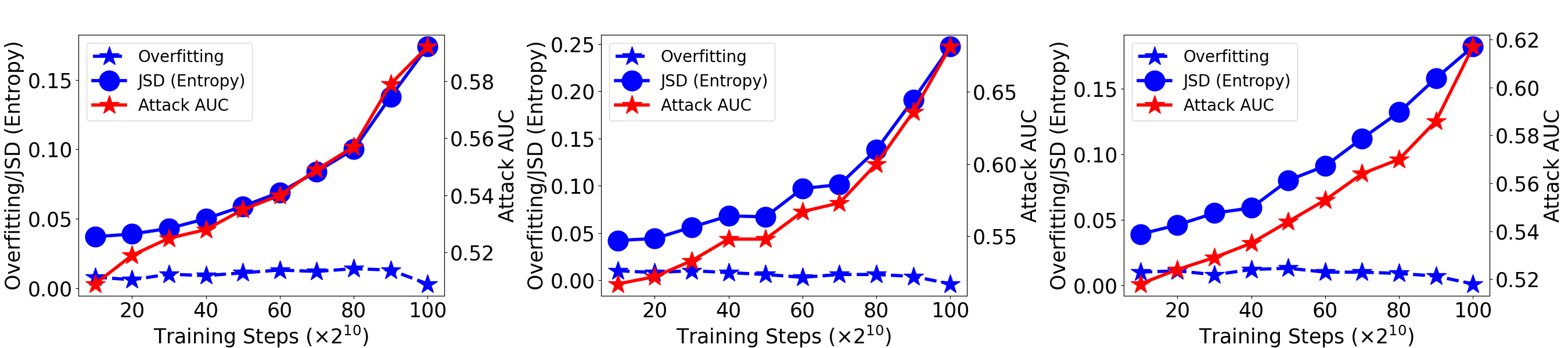}
\caption{SVHN}
\label{figure:combine_svhn_WideResNet_500_epoch_performance}
\end{subfigure}
\begin{subfigure}{\columnwidth}
\includegraphics[width=\columnwidth]{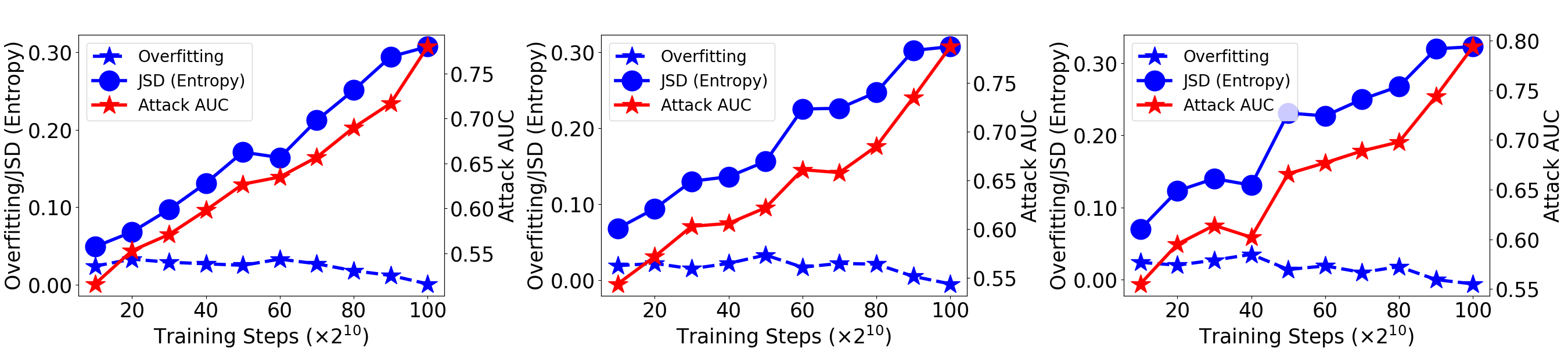}
\caption{CIFAR10}
\label{figure:combine_cifar10_WideResNet_500_epoch_performance}
\end{subfigure}
\begin{subfigure}{\columnwidth}
\includegraphics[width=\columnwidth]{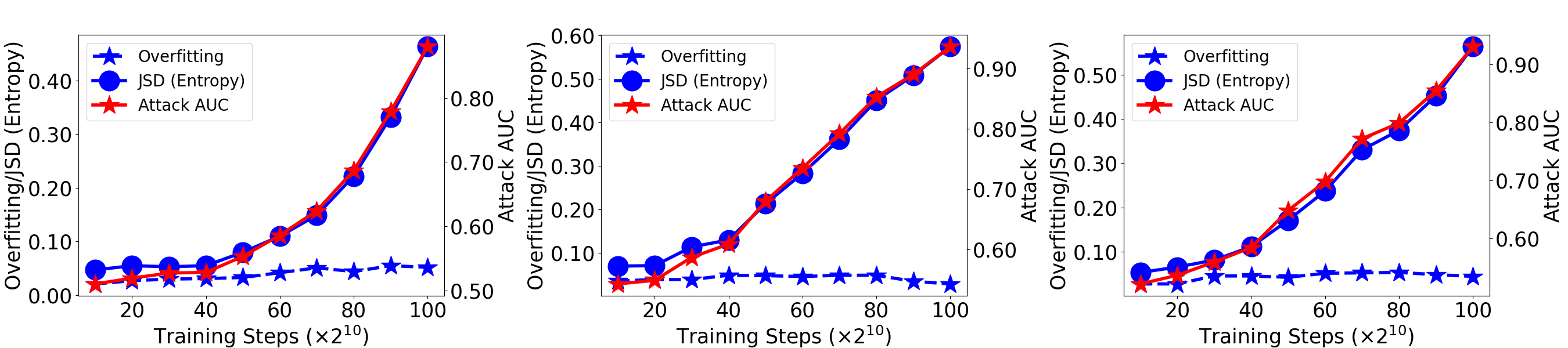}
\caption{CIFAR100}
\label{figure:combine_cifar100_WideResNet_500_epoch_performance}
\end{subfigure}
\caption{The overfitting/JS Distance (Entropy) and attack AUC with respect to different training steps. 
The first to third columns denotes the model trained by FixMatch, FlexMatch, and UDA with 500 labeled samples, respectively. 
Note that we consider the attack AUC of $\AttackModel_{DA}$, which is the strongest attack.}
\label{figure:epoch_performance_500}
\end{figure}

\begin{figure}[!t]
\centering
\begin{subfigure}{\columnwidth}
\includegraphics[width=\columnwidth]{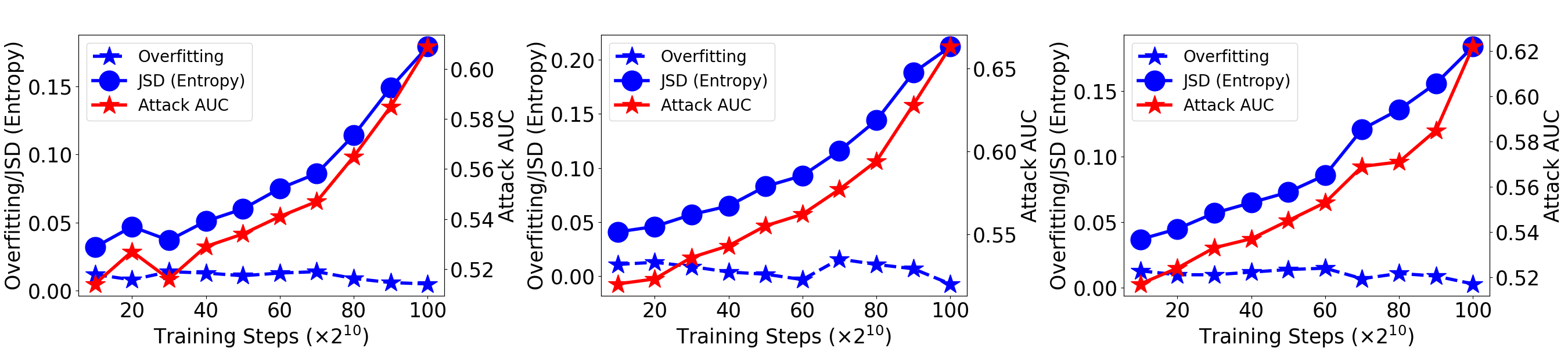}
\caption{SVHN}
\label{figure:combine_svhn_WideResNet_1000_epoch_performance}
\end{subfigure}
\begin{subfigure}{\columnwidth}
\includegraphics[width=\columnwidth]{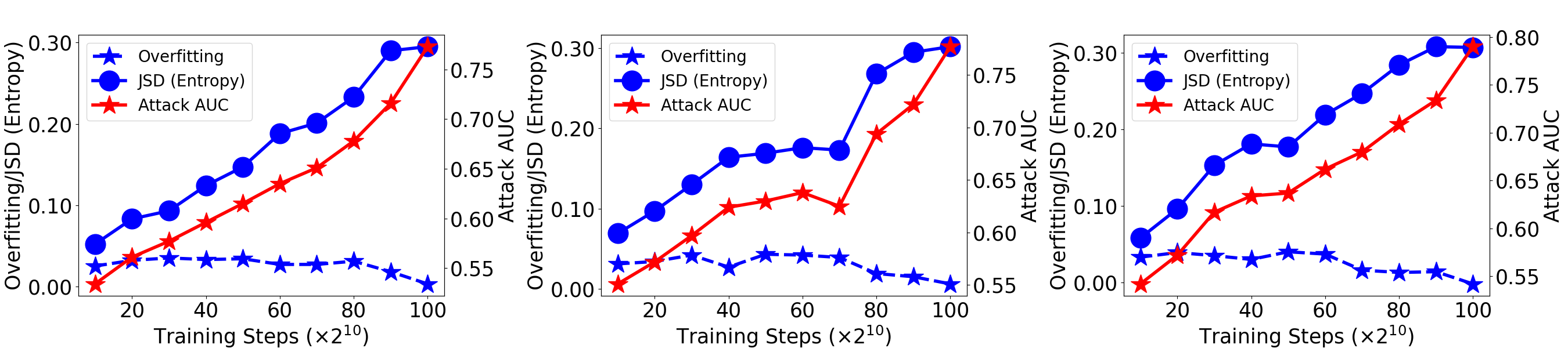}
\caption{CIFAR10}
\label{figure:combine_cifar10_WideResNet_1000_epoch_performance}
\end{subfigure}
\begin{subfigure}{\columnwidth}
\includegraphics[width=\columnwidth]{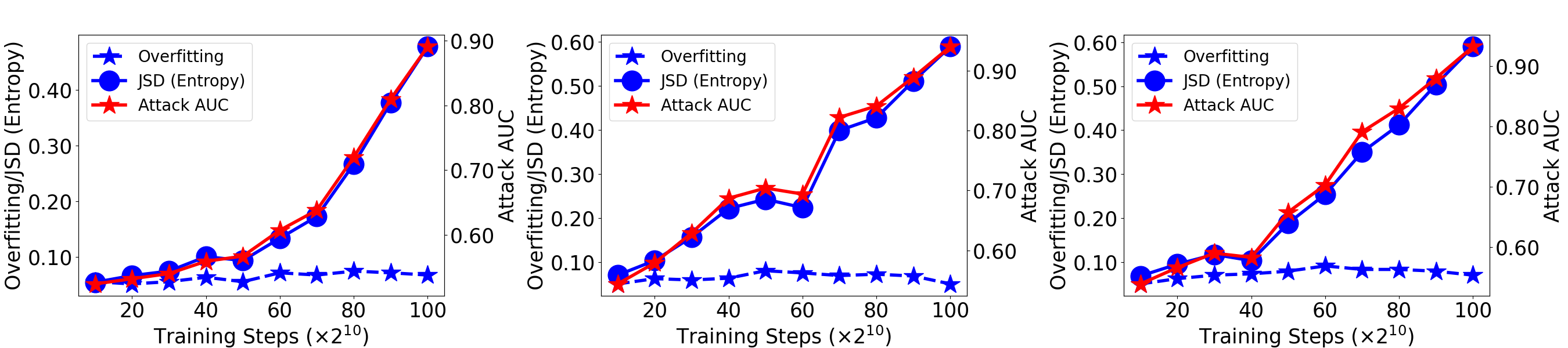}
\caption{CIFAR100}
\label{figure:combine_cifar100_WideResNet_1000_epoch_performance}
\end{subfigure}
\caption{The overfitting/JS Distance (Entropy) and attack AUC with respect to different training steps. 
The first to third columns denotes the model trained by FixMatch, FlexMatch, and UDA with 1,000 labeled samples, respectively. 
Note that we consider the attack AUC of $\AttackModel_{DA}$, which is the strongest attack.}
\label{figure:epoch_performance_1000}
\end{figure}

\begin{figure}[!t]
\centering
\begin{subfigure}{\columnwidth}
\includegraphics[width=\columnwidth]{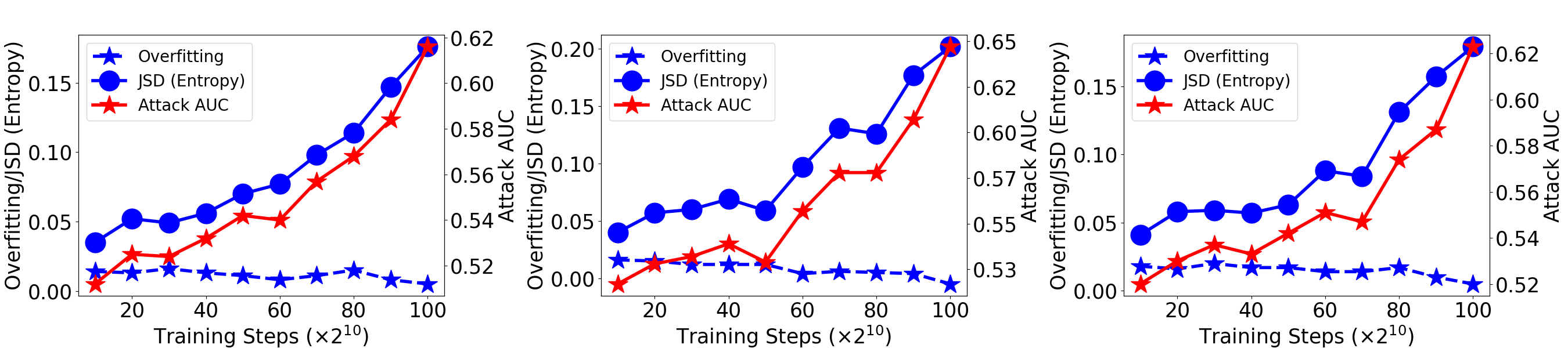}
\caption{SVHN}
\label{figure:combine_svhn_WideResNet_2000_epoch_performance}
\end{subfigure}
\begin{subfigure}{\columnwidth}
\includegraphics[width=\columnwidth]{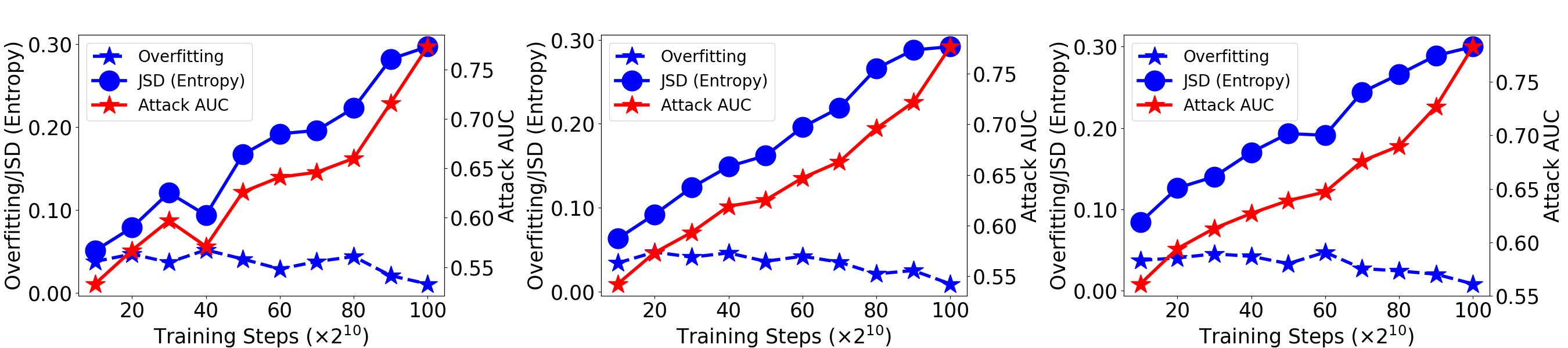}
\caption{CIFAR10}
\label{figure:combine_cifar10_WideResNet_2000_epoch_performance}
\end{subfigure}
\begin{subfigure}{\columnwidth}
\includegraphics[width=\columnwidth]{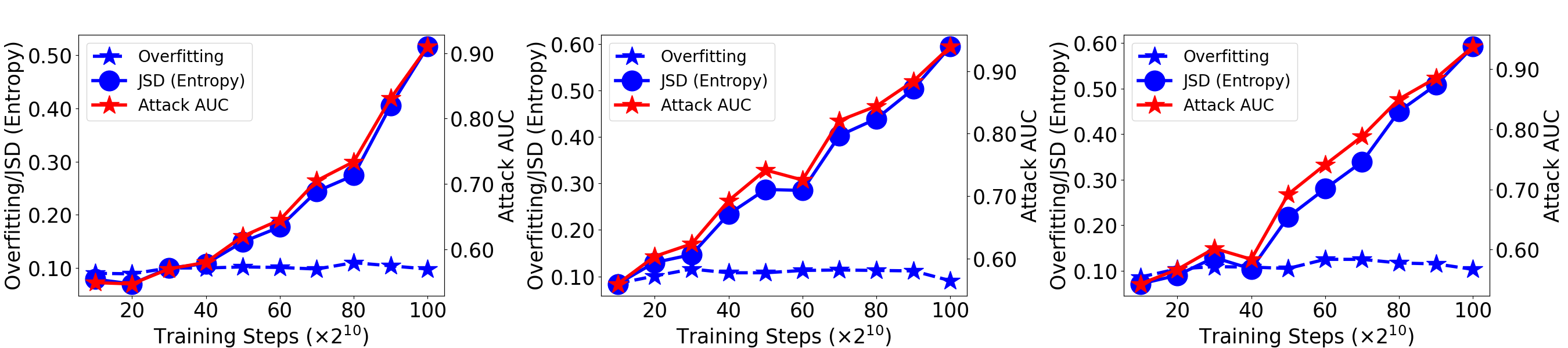}
\caption{CIFAR100}
\label{figure:combine_cifar100_WideResNet_2000_epoch_performance}
\end{subfigure}
\caption{The overfitting/JS Distance (Entropy) and attack AUC with respect to different training steps. 
The first to third columns denotes the model trained by FixMatch, FlexMatch, and UDA with 2,000 labeled samples, respectively. 
Note that we consider the attack AUC of $\AttackModel_{DA}$, which is the strongest attack.}
\label{figure:epoch_performance_2000}
\end{figure}

\begin{figure}[!t]
\centering
\begin{subfigure}{\columnwidth}
\includegraphics[width=\columnwidth]{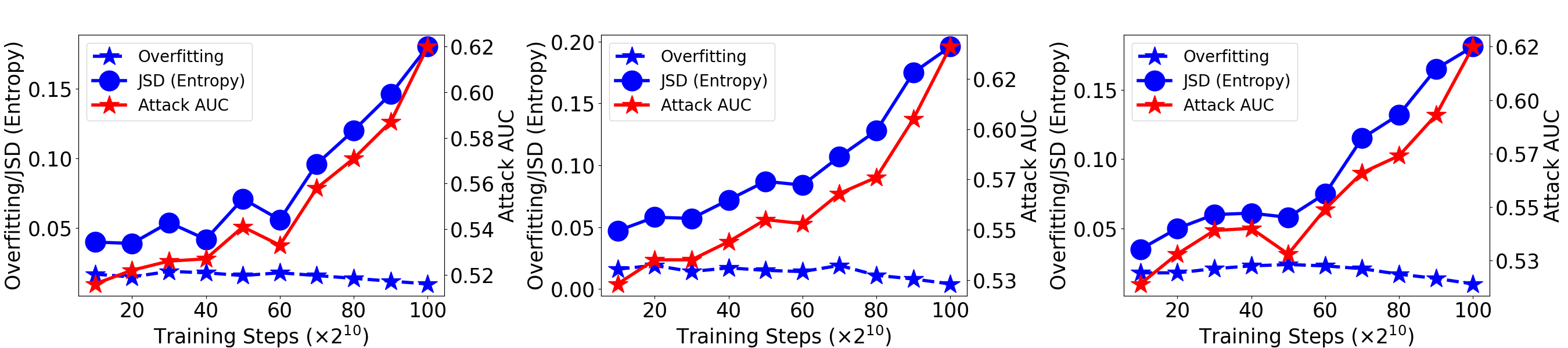}
\caption{SVHN}
\label{figure:combine_svhn_WideResNet_4000_epoch_performance}
\end{subfigure}
\begin{subfigure}{\columnwidth}
\includegraphics[width=\columnwidth]{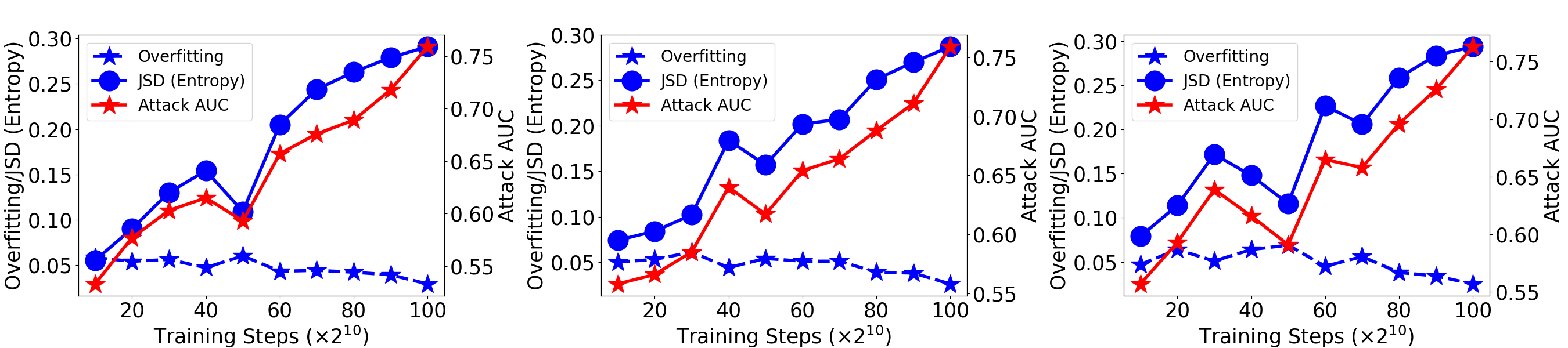}
\caption{CIFAR10}
\label{figure:combine_cifar10_WideResNet_4000_epoch_performance}
\end{subfigure}
\begin{subfigure}{\columnwidth}
\includegraphics[width=\columnwidth]{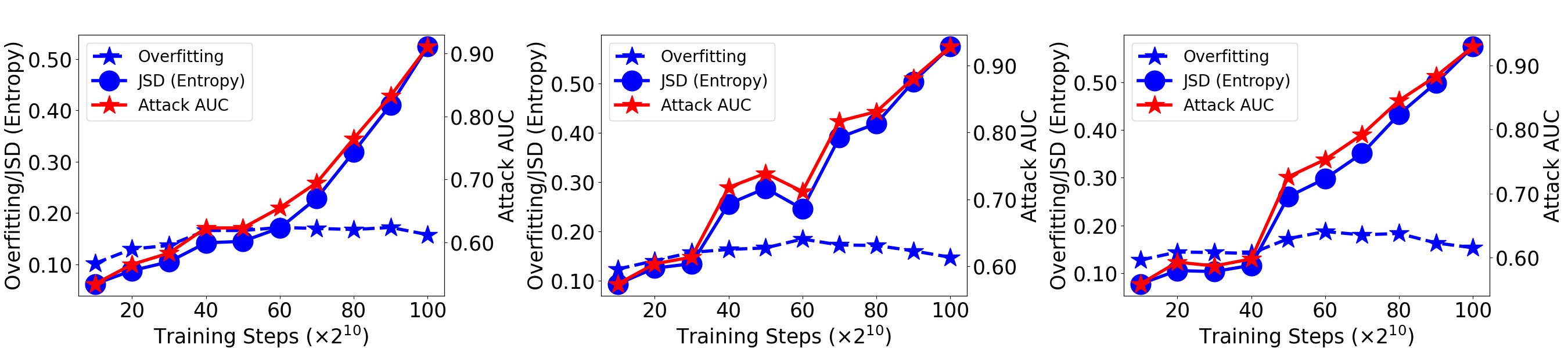}
\caption{CIFAR100}
\label{figure:combine_cifar100_WideResNet_4000_epoch_performance}
\end{subfigure}
\caption{The overfitting/JS Distance (Entropy) and attack AUC with respect to different training steps. 
The first to third columns denote the model trained by FixMatch, FlexMatch, and UDA with 4,000 labeled samples, respectively. 
Note that we consider the attack AUC of $\AttackModel_{DA}$, which is the strongest attack.}
\label{figure:epoch_performance_4000}
\end{figure}

\autoref{figure:epoch_performance_500}, \autoref{figure:epoch_performance_1000}, \autoref{figure:epoch_performance_2000}, and \autoref{figure:epoch_performance_4000} shows the results of models trained by different SSL methods on the three datasets with 500, 1,000, 2,000, and 4,000 labeled samples.
We has the similar finding as \autoref{subsection:epoch_wised_performance}, i.e., the models trained by SSL has almost no overfitting, but the JS Distance (Entropy) and the attack performance do increase during the training.

% ----------------------------------------------------
\subsection{Ablation Study: Number of Views}
\label{appendix:ablation_number_of_view}
% ----------------------------------------------------

\begin{figure}[!t]
\centering
\begin{subfigure}{0.3\columnwidth}
\includegraphics[width=\columnwidth]{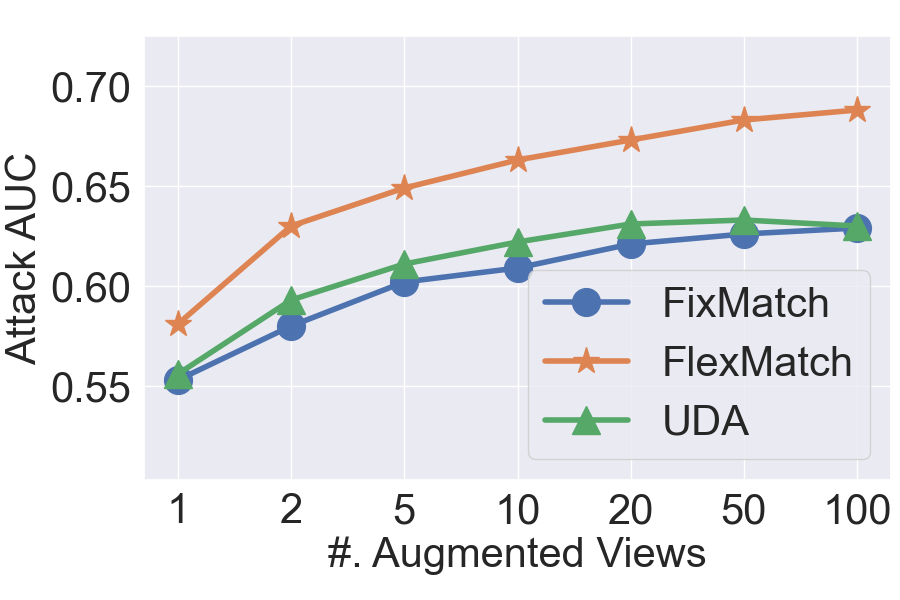}
\caption{SVHN}
\label{figure:svhn_1000_ablation_view_test_auc}
\end{subfigure}
\begin{subfigure}{0.3\columnwidth}
\includegraphics[width=\columnwidth]{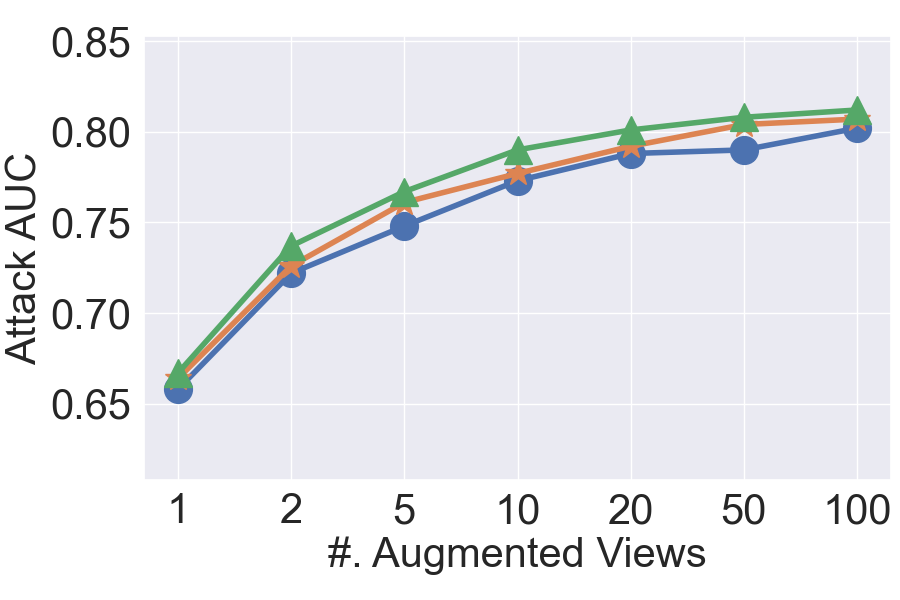}
\caption{CIFAR10}
\label{figure:cifar10_1000_ablation_view_test_auc}
\end{subfigure}
\begin{subfigure}{0.3\columnwidth}
\includegraphics[width=\columnwidth]{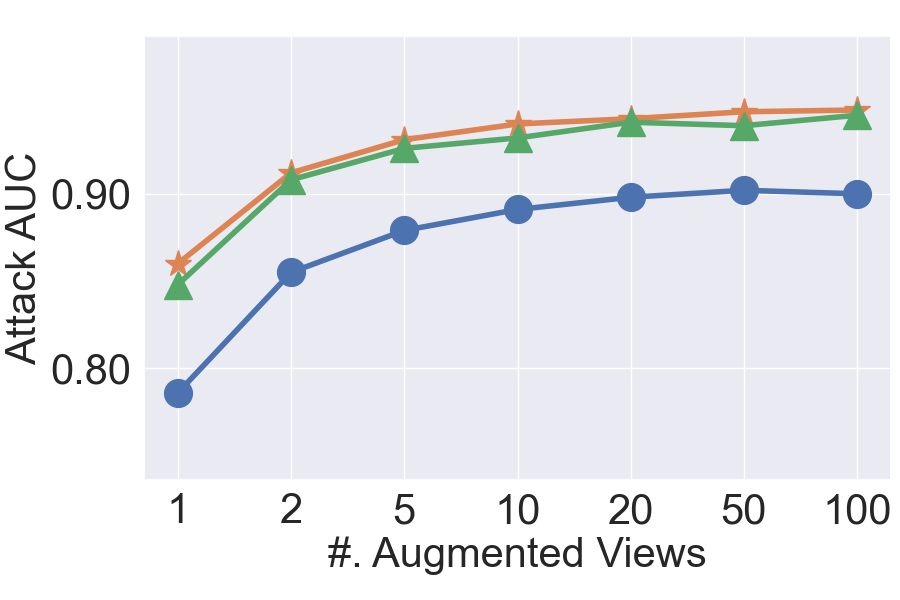}
\caption{CIFAR100}
\label{figure:cifar100_1000_ablation_view_test_auc}
\end{subfigure}
\caption{The attack AUC of $\AttackModel_{DA}$  with different numbers of augmented views to query the target model. 
The target model is trained with 1,000 labeled samples.}
\label{figure:1000_ablation_view_test_auc}
\end{figure}

\begin{figure}[!t]
\centering
\begin{subfigure}{0.3\columnwidth}
\includegraphics[width=\columnwidth]{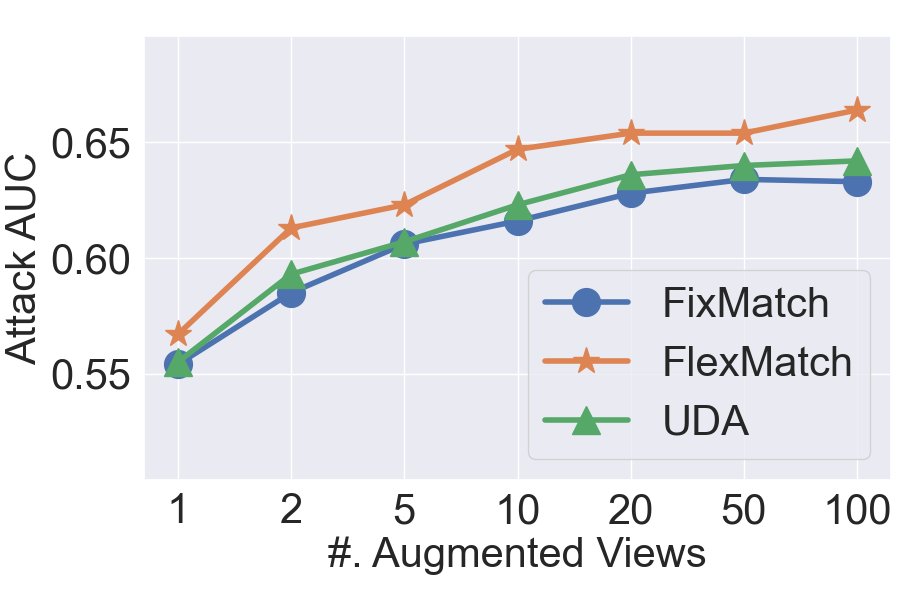}
\caption{SVHN}
\label{figure:svhn_2000_ablation_view_test_auc}
\end{subfigure}
\begin{subfigure}{0.3\columnwidth}
\includegraphics[width=\columnwidth]{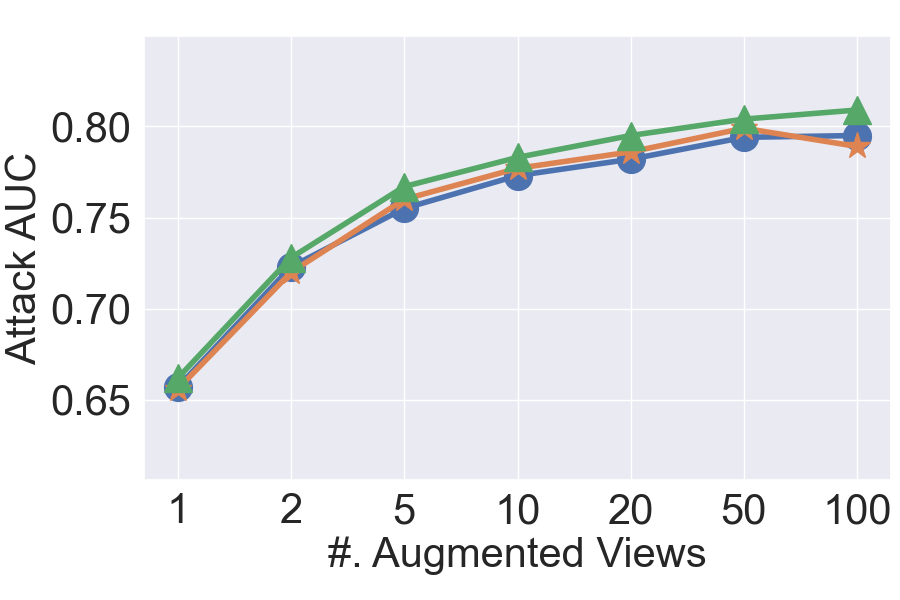}
\caption{CIFAR10}
\label{figure:cifar10_2000_ablation_view_test_auc}
\end{subfigure}
\begin{subfigure}{0.3\columnwidth}
\includegraphics[width=\columnwidth]{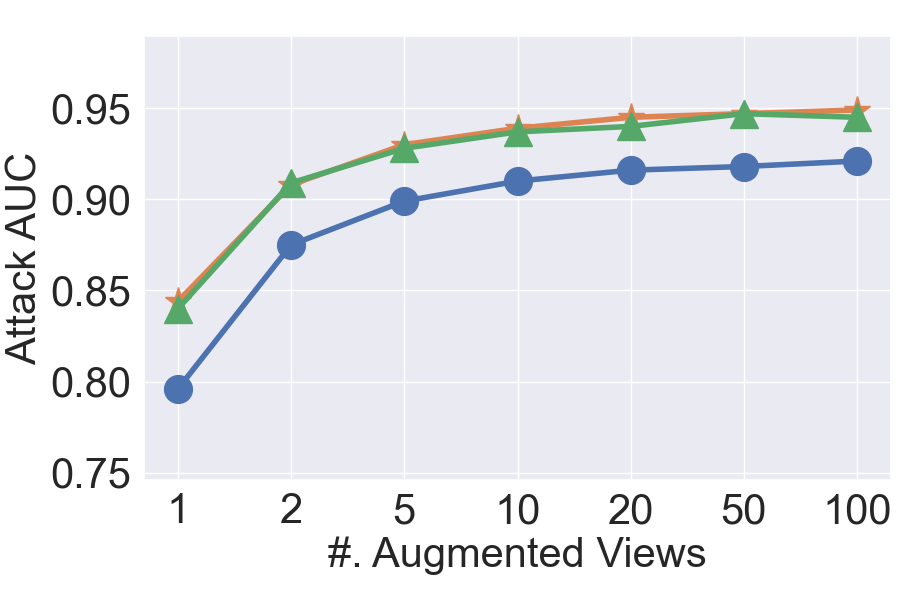}
\caption{CIFAR100}
\label{figure:cifar100_2000_ablation_view_test_auc}
\end{subfigure}
\caption{The attack AUC of $\AttackModel_{DA}$ with different numbers of augmented views to query the target model. 
The target model is trained with 2,000 labeled samples.}
\label{figure:2000_ablation_view_test_auc}
\end{figure}

\begin{figure}[!t]
\centering
\begin{subfigure}{0.3\columnwidth}
\includegraphics[width=\columnwidth]{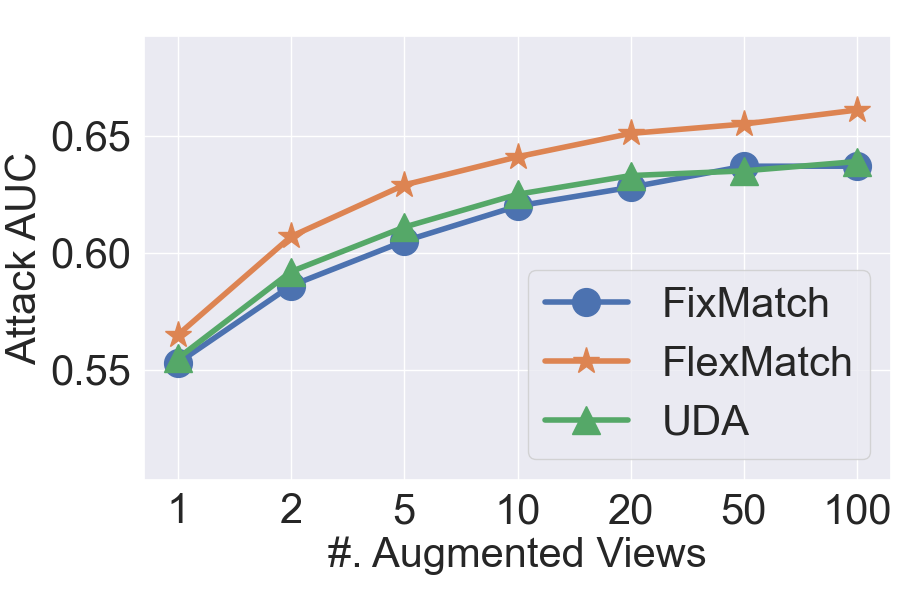}
\caption{SVHN}
\label{figure:svhn_4000_ablation_view_test_auc}
\end{subfigure}
\begin{subfigure}{0.3\columnwidth}
\includegraphics[width=\columnwidth]{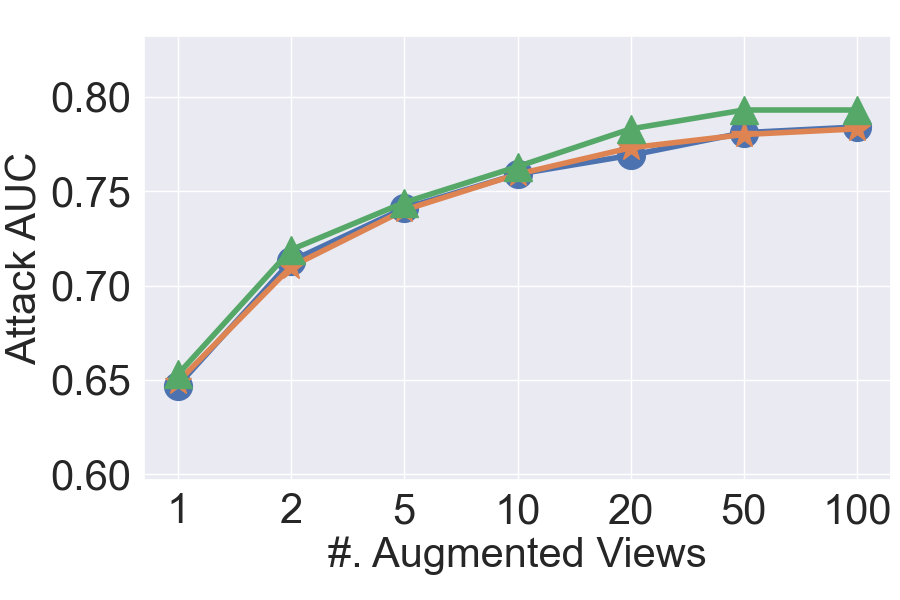}
\caption{CIFAR10}
\label{figure:cifar10_4000_ablation_view_test_auc}
\end{subfigure}
\begin{subfigure}{0.3\columnwidth}
\includegraphics[width=\columnwidth]{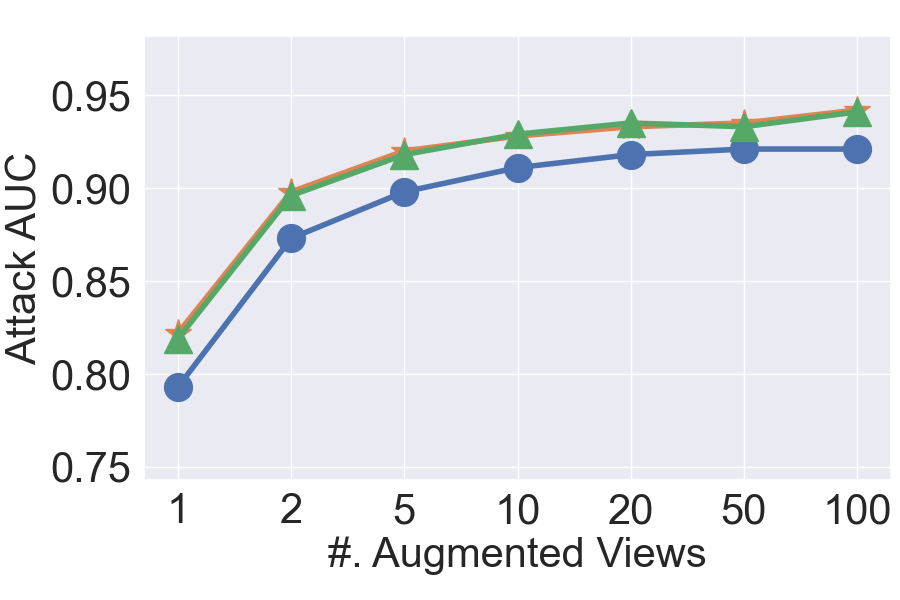}
\caption{CIFAR100}
\label{figure:cifar100_4000_ablation_view_test_auc}
\end{subfigure}
\caption{The attack AUC of $\AttackModel_{DA}$  with different numbers of augmented views to query the target model. 
The target model is trained with 4,000 labeled samples.}
\label{figure:4000_ablation_view_test_auc}
\end{figure}

For the SSL methods trained on different datasets with 1,000, 2,000, and 4,000 labeled samples, we range the number of views from 1 to 100 and the attack performance is shown in
\autoref{figure:1000_ablation_view_test_auc}, \autoref{figure:2000_ablation_view_test_auc}, and \autoref{figure:4000_ablation_view_test_auc}, respectively.
We have the similar observation as \autoref{subsection:ablation_study_attack_model} that more number of views leads to better performance.

% ----------------------------------------------------
\subsection{Ablation Study: Similarity Function}
\label{appendix:ablation_similarity_function}
% ----------------------------------------------------

\begin{figure}[!t]
\centering
\begin{subfigure}{0.3\columnwidth}
\includegraphics[width=\columnwidth]{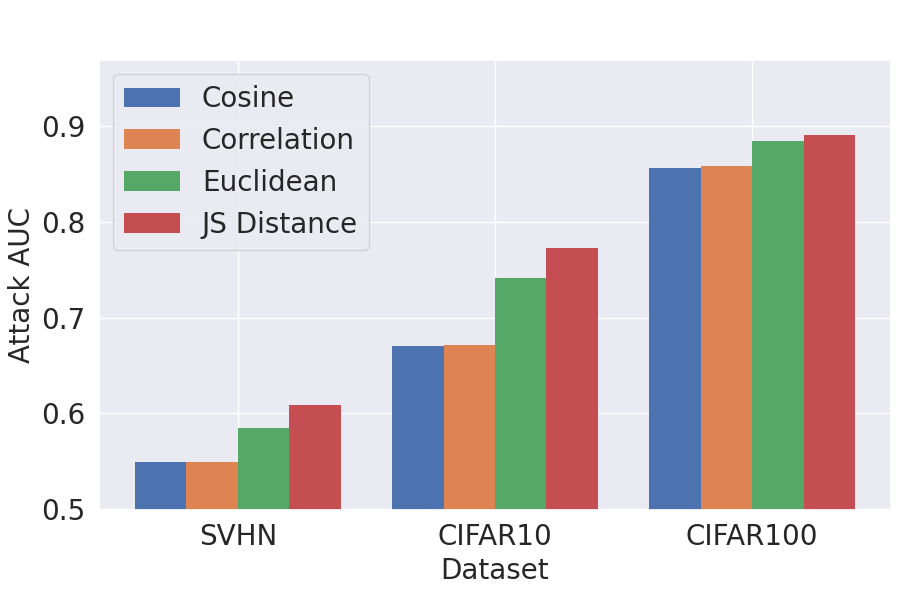}
\caption{FixMatch}
\label{figure:fixmatch_1000_ablation_similarity_func_test_auc}
\end{subfigure}
\begin{subfigure}{0.3\columnwidth}
\includegraphics[width=\columnwidth]{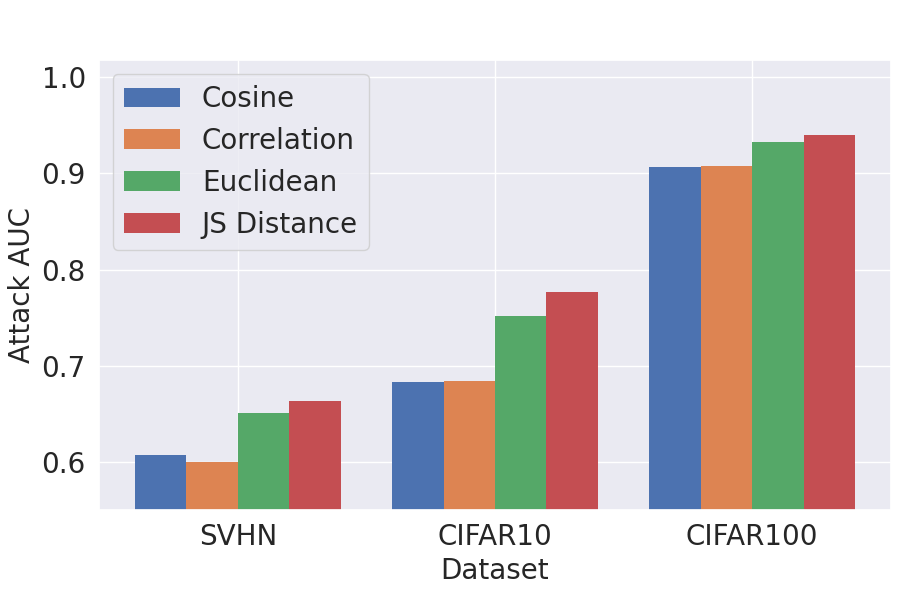}
\caption{FlexMatch}
\label{figure:flexmatch_1000_ablation_similarity_func_test_auc}
\end{subfigure}
\begin{subfigure}{0.3\columnwidth}
\includegraphics[width=\columnwidth]{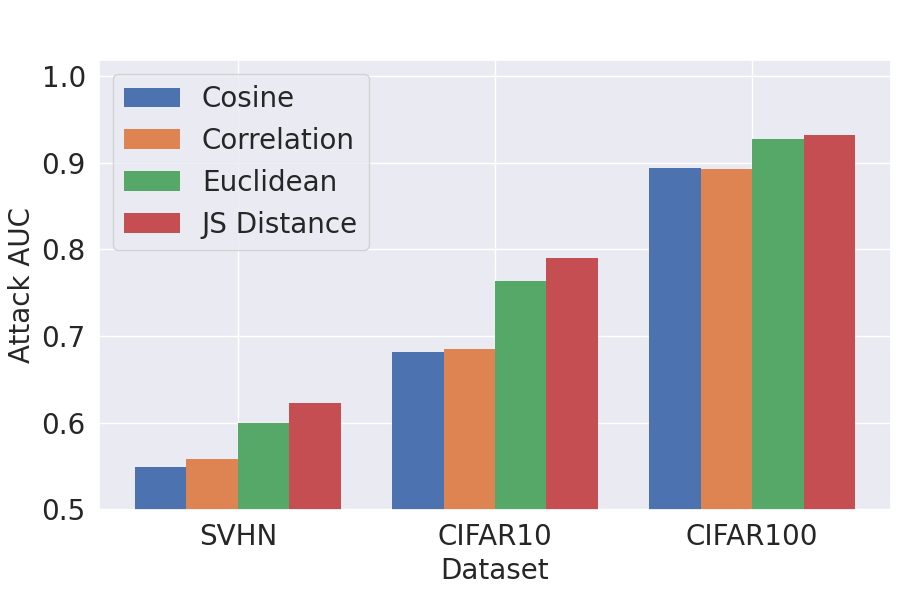}
\caption{UDA}
\label{figure:uda_1000_ablation_similarity_func_test_auc}
\end{subfigure}
\caption{The attack AUC of $\AttackModel_{DA}$  with different similarity functions. 
The target model is trained with 1,000 labeled samples.}
\label{figure:ablation_similarity_func_test_auc_1000}
\end{figure}

\begin{figure}[!t]
\centering
\begin{subfigure}{0.3\columnwidth}
\includegraphics[width=\columnwidth]{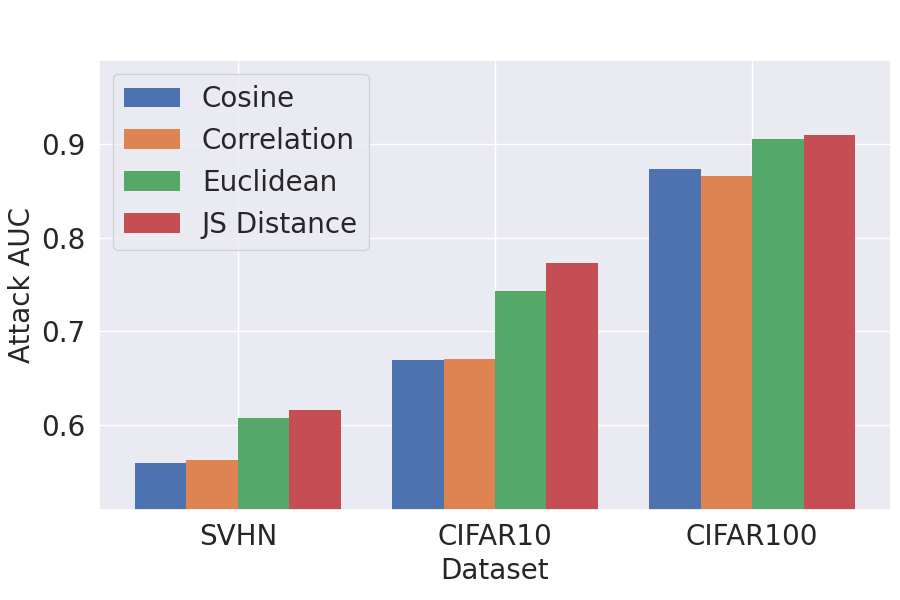}
\caption{FixMatch}
\label{figure:fixmatch_2000_ablation_similarity_func_test_auc}
\end{subfigure}
\begin{subfigure}{0.3\columnwidth}
\includegraphics[width=\columnwidth]{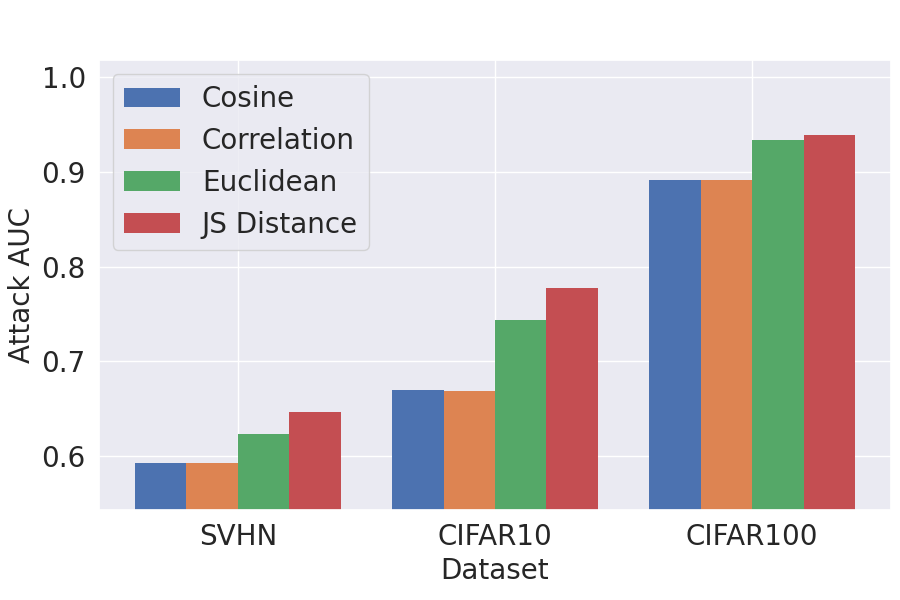}
\caption{FlexMatch}
\label{figure:flexmatch_2000_ablation_similarity_func_test_auc}
\end{subfigure}
\begin{subfigure}{0.3\columnwidth}
\includegraphics[width=\columnwidth]{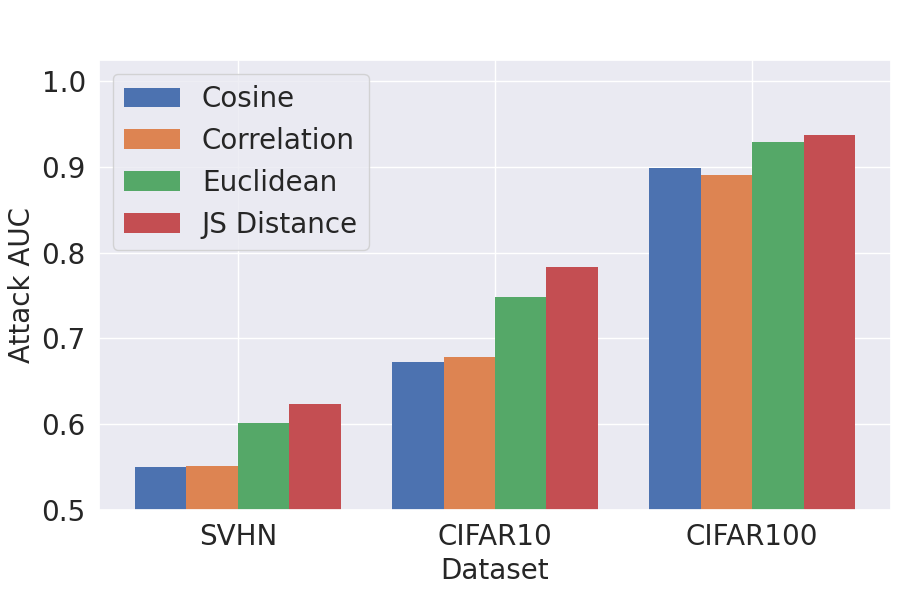}
\caption{UDA}
\label{figure:uda_2000_ablation_similarity_func_test_auc}
\end{subfigure}
\caption{The attack AUC of $\AttackModel_{DA}$  with different similarity functions. 
The target model is trained with 2,000 labeled samples.}
\label{figure:ablation_similarity_func_test_auc_2000}
\end{figure}

\begin{figure}[!t]
\centering
\begin{subfigure}{0.3\columnwidth}
\includegraphics[width=\columnwidth]{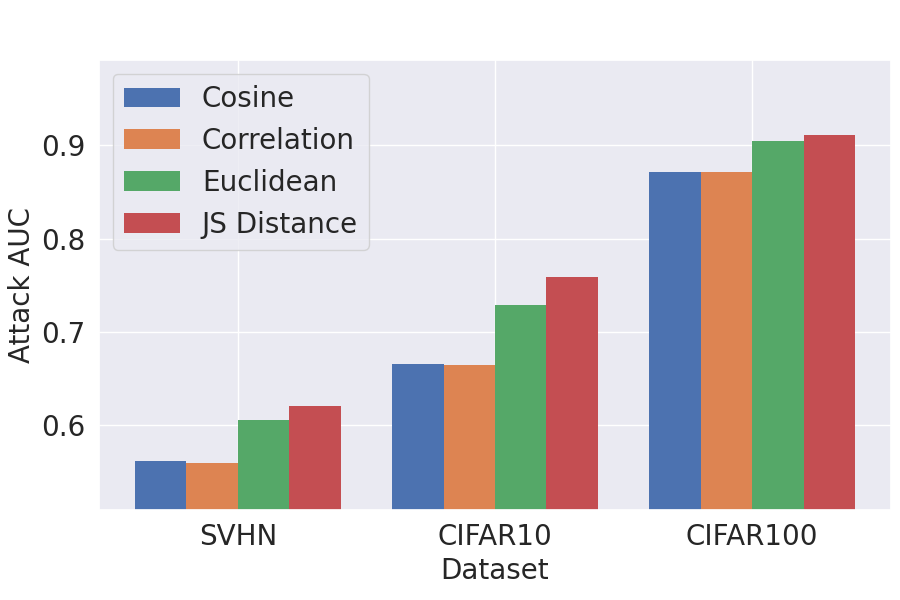}
\caption{FixMatch}
\label{figure:fixmatch_4000_ablation_similarity_func_test_auc}
\end{subfigure}
\begin{subfigure}{0.3\columnwidth}
\includegraphics[width=\columnwidth]{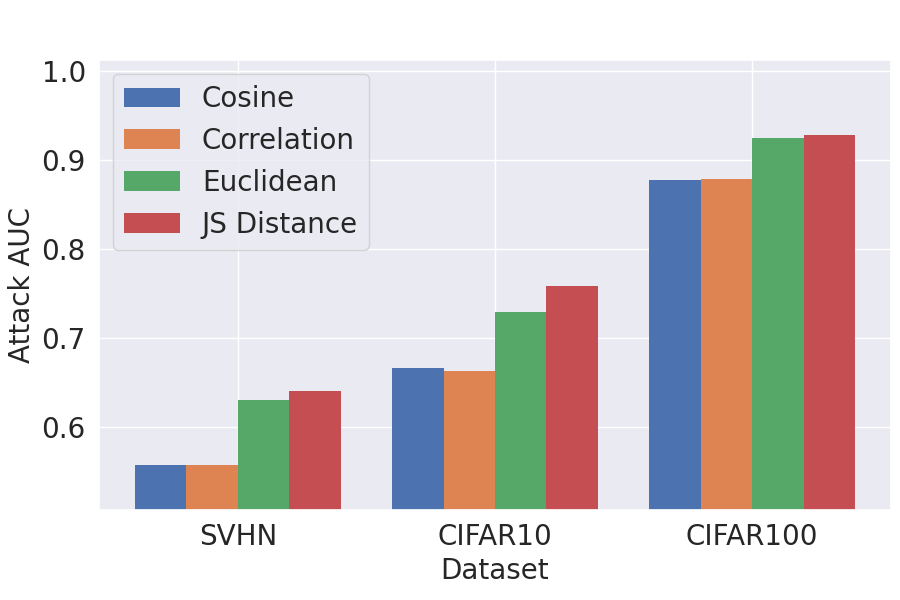}
\caption{FlexMatch}
\label{figure:flexmatch_4000_ablation_similarity_func_test_auc}
\end{subfigure}
\begin{subfigure}{0.3\columnwidth}
\includegraphics[width=\columnwidth]{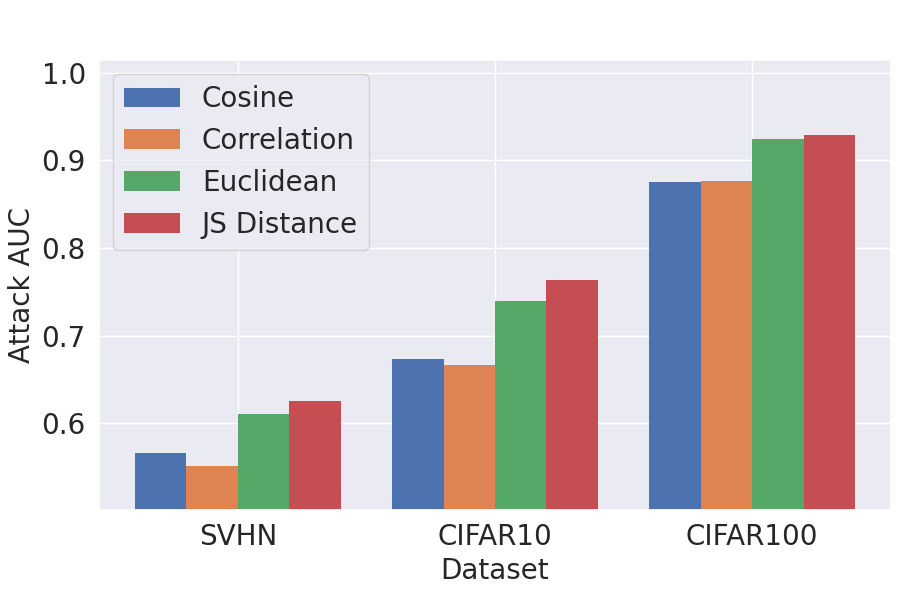}
\caption{UDA}
\label{figure:uda_4000_ablation_similarity_func_test_auc}
\end{subfigure}
\caption{The attack AUC of $\AttackModel_{DA}$  with different similarity functions. 
The target model is trained with 4,000 labeled samples.}
\label{figure:ablation_similarity_func_test_auc_4000}
\end{figure}

The results for FixMatch, FlexMatch, and UDA trained on different datasets with 1,000, 2,000, and 4,000 labeled samples are shown in \autoref{figure:ablation_similarity_func_test_auc_1000}, \autoref{figure:ablation_similarity_func_test_auc_2000},
and \autoref{figure:ablation_similarity_func_test_auc_4000}, respectively.
We observe that the JS Distance still consistently outperforms the other three distance metrics and achieves the best performance.

% ----------------------------------------------------
\subsection{Ablation Study: Data Augmentation and Shadow Model Architecture}
\label{appendix:ablation_attack_model}
% ----------------------------------------------------

\mypara{Data Augmentation}
In the previous evaluation of $\AttackModel_{DA}$, we assume the adversary knows the data augmentation used to train the target model and can apply the same data augmentation to conduct the attack.
We then relax this assumption to see whether $\AttackModel_{DA}$ is still effective with different levels of data augmentations.
We take FixMatch trained on CIFAR10 with 500 labeled samples as a case study amd the results are shown in \autoref{table:ablation_augmentation}.
We find that $\AttackModel_{DA}$ is still effective even with the weakest augmentation (Aug-Level is 0) and the attack performance can be improved with stronger augmentations added.
For instance, the attack AUC is 0.876 and 0.882 when the Aug-Level is 0 and 4, respectively.
This implies that a successful attack can still be launched even without the exact knowledge of the data augmentation used to train the target model.

\begin{table*}[]
\caption{The attack performance with respect to different augmentation levels.
For Aug-Level=0, we only apply random cropping with flipping as both weak and strong augmentation.
For 1-4, we gradually apply 1-4 transformation methods for the strong augmentation to each image from the general augmentation pools. 
The target model is trained by FixMatch on CIFAR100 with 500 labeled samples. ($\star$) denotes our default setting in the paper.}
\label{table:ablation_augmentation}
\centering
\begin{tabular}{l | c |c |c | c | c}
\toprule
Aug-Level & 0 & 1 & 2 ($\star$) & 3 & 4 \\
\midrule
Attack AUC & 0.876 & 0.880 & 0.880 & 0.881 & 0.882\\
\bottomrule
\end{tabular}
\end{table*}

\mypara{Shadow Model Architecture}
We then relax the assumption that the shadow model has the same architecture as the target model.
Given the target model (WRN28-2 trained by FixMatch on CIFAR100 with 500 labeled samples), we train shadow models with WRN28-2, WRN28-1, WRN28-4, WRN28-8, and ResNet50 as the architectures, and the corresponding attack AUCs are 0.880, 0.875, 0.839, 0.835, and 0.859, respectively.
Our attack achieves the highest attack AUC when the shadow and target models use the same architecture. 
However, our attack is still effective even if the shadow model has a different model architecture (e.g., the attack AUC is 0.859 when the shadow model architecture is ResNet50).

% ----------------------------------------------------
\subsection{Defense Evaluation}
\label{appendix:defense}
% ----------------------------------------------------

Besides early stopping (ES), we evaluate three more defenses, i.e., top-$k$ posteriors (top-$k$)~\cite{SSSS17}, model stacking (MS)~\cite{SZHBFB19}, and DP-SGD~\cite{ACGMMTZ16}. 
We evaluate both the target model's utility (test acc) and the effectiveness of defenses (attack auc).
The results are shown in \autoref{table:defense}.
We observe that DP-SGD is the most effective defense since it achieves the lowest attack AUC with 0.598.
However, DP-SGD suffers from unacceptable utility drop (with only 0.034 test acc).
Existing work~\cite{JE19,LLR21} on DP also show that DP-SGD sacrifices the model's utility substantially in order to achieve good privacy guarantee.
Therefore, we consider early stopping as the best defense since it achieves the best privacy-utility trade-off. i.e., it reduces the membership leakage to a large extent (from 0.918 to 0.695) while maintaining the utility (from 0.530 to 0.490).

\begin{figure}[!t]
\centering
\begin{subfigure}{0.3\columnwidth}
\includegraphics[width=\columnwidth]{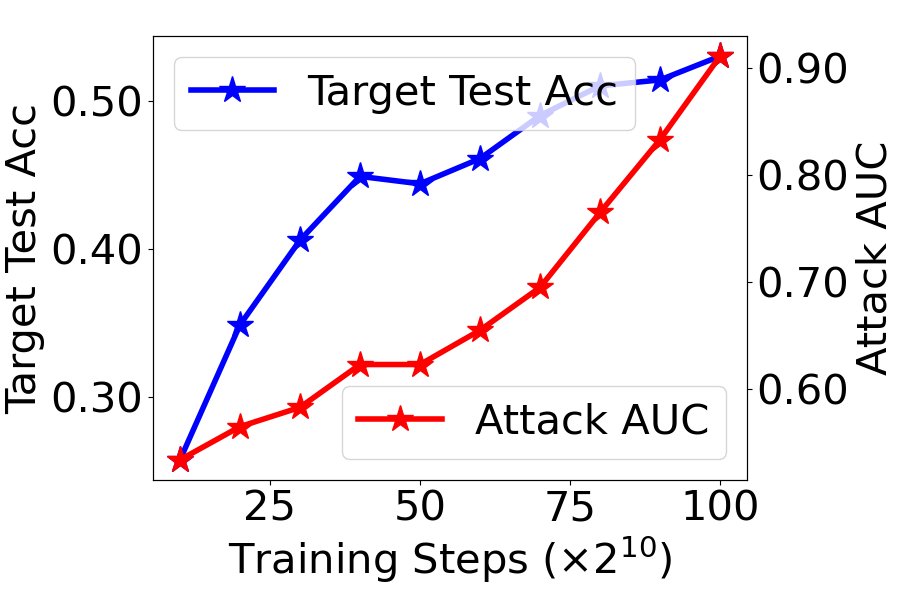}
\caption{FixMatch}
\label{figure:defense_earlystopping_cifar100_4000_fixmatch}
\end{subfigure}
\begin{subfigure}{0.3\columnwidth}
\includegraphics[width=\columnwidth]{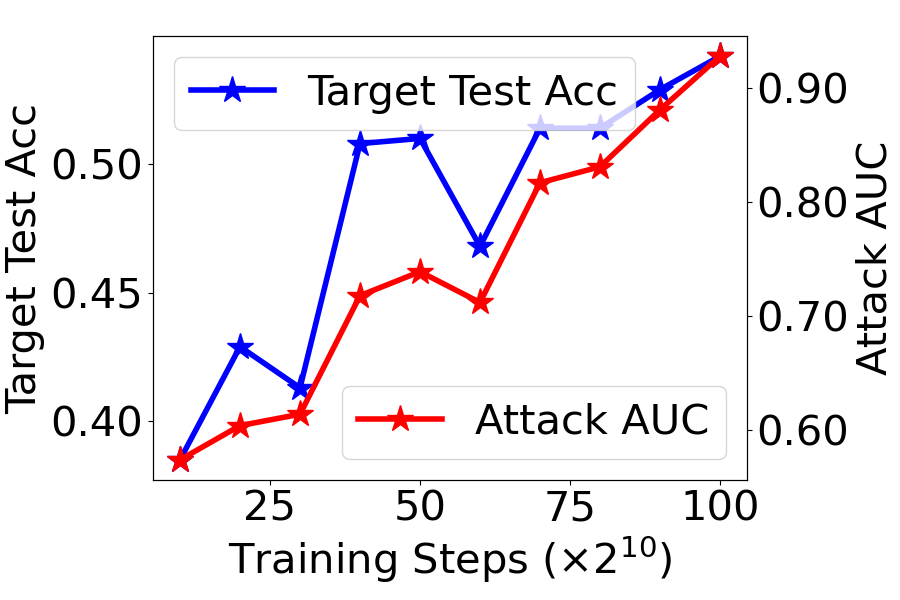}
\caption{FlexMatch}
\label{figure:defense_earlystopping_cifar100_4000_flexmatch}
\end{subfigure}
\begin{subfigure}{0.3\columnwidth}
\includegraphics[width=\columnwidth]{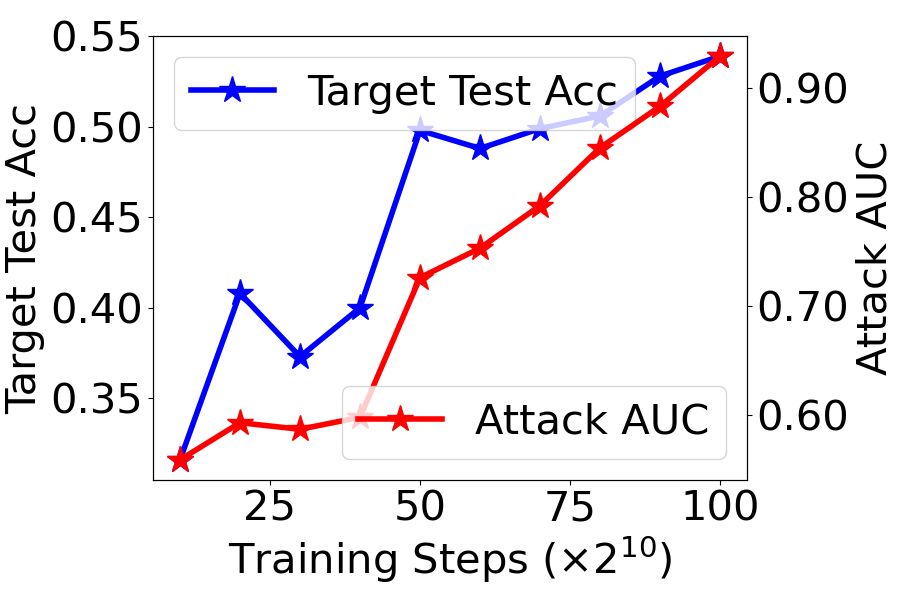}
\caption{UDA}
\label{figure:defense_earlystopping_cifar100_4000_uda}
\end{subfigure}
\caption{The target model performance and attack AUC with respect to different training steps. 
The target model is trained on CIFAR100 with 4,000 labeled samples, which has the highest performance on its original classification task. 
The attack model is $\AttackModel_{DA}$, which has the best attack performance.}
\label{figure:defense_earlystopping_cifar100_4000}
\end{figure}

\begin{table}[]
\caption{Target model accuracy and attack AUC for different defenses. 
The target model is WRN28-2 trained by FixMatch on CIFAR100 with 4,000 labeled samples (same setting as the paper). 
For early stopping (ES), we stop at 70 $\times 2^{10}$ training steps. 
For top-$k$, we set $k$=1 as it leaks the least information. For model stacking (MS), we train four models, i.e., WRN28-\{1,2,4,8\}, using the same dataset and average their posteriors. 
For DP-SGD, the noise scale is set to $10^{-5}$ and the gradient norm is set to 1. 
Note that we use the Opacus library~\cite{OPACUS} to implement DP-SGD.}
\label{table:defense}
\centering
\begin{tabular}{l | c | c | c | c | c}
\toprule
Defense & None & ES & Top-$k$ & MS & DP-SGD \\
\midrule
Test ACC & 0.530 & 0.490 & 0.530  & 0.549 & 0.034 \\
Attack AUC & 0.918 & 0.695 & 0.906 & 0.905 & 0.598 \\
\bottomrule
\end{tabular}
\end{table}

\setlength{\algomargin}{0em}
\begin{algorithm}[!ht]
    \caption{Our Data Augmentation Based Attack $\AttackModel_{DA}$}
    \begin{algorithmic}
    \STATE {\bfseries Input:} Given sample $x$, target model $\TargetModel$, NN-based attack model $\AttackModel_{DA}$, weak data augmentations $\mathcal{A}ug_{weak}$, strong data augmentations $\mathcal{A}ug_{strong}$, similarity function $Sim$, number of augmented views $K$\\
    \STATE {\bfseries Output:} Member or non-member \\
    \tcc{Generate augmented views.}
    \STATE $\{x_{weak}^1, x_{weak}^2, \cdots, x_{weak}^K\} \leftarrow \mathcal{A}ug_{weak}(x,K)$
    \STATE $\{x_{strong}^1, x_{strong}^2, \cdots, x_{strong}^K\} \leftarrow \mathcal{A}ug_{strong}(x,K)$
    
    \tcc{Query the target model and obtain posteriors.}
    \STATE $\{p_{weak}^1, p_{weak}^2, \cdots, p_{weak}^K\} \leftarrow \{\TargetModel(x_{weak}^1), \TargetModel(x_{weak}^2), \cdots, \TargetModel{x_{weak}^K}\}$
    \STATE $\{p_{strong}^1, p_{strong}^2, \cdots, p_{strong}^K\} \leftarrow \{\TargetModel(x_{strong}^1), \TargetModel(x_{strong}^2), \cdots, \TargetModel{x_{strong}^K}\}$
    
    \tcc{Obtain similarity vectors.}
    \STATE Similarity vector $v_{w}(x) \leftarrow \operatorname{SORTED}(\{Sim(p_{weak}^i, p_{weak}^j) | i\in [1,K], j \in [1,K]  \})$
    \STATE Similarity vector $v_{s}(x) \leftarrow \operatorname{SORTED}(\{Sim(p_{strong}^i, p_{strong}^j) | i\in [1,K], j \in [1,K]  \})$
    \STATE Similarity vector $v_{ws}(x) \leftarrow \operatorname{SORTED}(\{Sim(p_{weak}^i, p_{strong}^j) | i\in [1,K], j \in [1,K]  \})$
    \tcc{Concatenate similarity vectors and perform the attack.}
    \STATE Merged vector $v(x)  \leftarrow \operatorname{CONCATENATE}(v_{w}(x),v_{s}(x),v_{ws}(x))$ 
    \STATE \textbf{return} $\AttackModel_{DA}(v(x))$\\
    \end{algorithmic}
\label{algorithml1}
\end{algorithm}

\end{document}